\documentclass{caps}
\usepackage{epsfig}
\usepackage{epsf}
\usepackage{aastable}
\usepackage{rotating}
\usepackage{morefloats} 

\catcode`@11
\def\x{@{\extracolsep{\fill}}}
\def\tabnoteentry#1#2{\par\noindent{\@hangfrom{$^{#1}$\hskip2pt}#2\par}}
\newenvironment{tabnote}{\par
\footnotesize
}{%
  \par}
\catcode`@12

\newcommand\lesssim{\mbox{$^{<}\hspace{-0.24cm}_{\sim}$}}
\newcommand\gtrsim{\mbox{$^{>}\hspace{-0.24cm}_{\sim}$}}
\def\msun{$M_{\odot}$}
\def\rsun{$R_{\odot}$}
\def\mdot{$\dot M$}
\def\eddmdot{$\dot m$}
\def\ergsec{\hbox{erg s$^{-1}$ }}

\def\ergcm{\hbox{erg cm$^{-2}$ s$^{-1}$ }}

\def\RS{$R_{\rm S}$}
\def\Rg{$R_{\rm g}$}

\begin{document}

\pagenumbering{roman}
\cleardoublepage
\pagenumbering{arabic}

\setcounter{chapter}{3}
\setcounter{table}{0}

\author{Jeffrey E. McClintock\\Harvard--Smithsonian Center
for Astrophysics, 60 Garden St., Cambridge, MA 02138, USA \and
Ronald A. Remillard\\Center for Space Research, MIT,
Cambridge, MA 02139, USA}

\chapter{Black Hole Binaries}

\title{Black Hole Binaries}

\section{Introduction}

\subsection{Scope of this review}

We focus on 18 black holes with measured masses that are located in
X--ray binary systems.  These black holes are the most visible
representatives of an estimated $\sim$300~million stellar--mass black
holes that are believed to exist in the Galaxy (van den Heuvel 1992;
Brown \& Bethe 1994; Timmes et al. 1996; Agol et al. 2002).  Thus the
mass of this particular form of dark matter, assuming
$\sim$10\msun~per black hole, is $\sim$4\% of the total baryonic mass
(i.e., stars plus gas) of the Galaxy (Bahcall 1986; Bronfman et
al. 1988).  Collectively this vast population of black holes outweighs
the Galactic--center black hole, SgrA$^{*}$, by a factor of
$\sim$1000.  These stellar--mass black holes are important to
astronomy in numerous ways.  For example, they are one endpoint of
stellar evolution for massive stars, and the collapse of their
progenitor stars enriches the universe with heavy elements (Woosley et
al. 2002).  Also, the measured mass distribution for even the small
sample of 18 black holes featured here are used to constrain models of
black hole formation and binary evolution (Brown et al. 2000a;
Nelemans \& van den Heuvel 2001; Fryer \& Kalogera 2001).  Lastly,
some black hole binaries appear to be linked to the hypernovae
believed to power gamma--ray bursts (Israelian et al. 1999; Brown et
al. 2000b; Orosz et al. 2001).

This review is focused on the X--ray timing and spectral properties of
these 18 black holes, plus a number of black hole candidates, with an
eye to their importance to physics as potential sites for tests of
general relativity (GR) in the strongest possible gravitational
fields.  There are now several current areas of research that probe
phenomena in these systems that are believed to occur very near the
event horizon.  These X--ray phenomena include quasi--periodic
oscillations (QPOs) at high frequency (40--450~Hz) observed from seven
systems, relativistically broadened iron lines from the inner
accretion disk, and thermal disk emission from near the innermost
stable circular orbit allowed by GR.  We also comment on evidence for
the existence of the event horizon, which is based on a comparison of
black--hole and neutron--star binaries and on models for advective
accretion flows.

The black hole binaries featured here are mass--exchange binaries that
contain an accreting black hole primary and a nondegenerate secondary
star.  They comprise about 10\% of all bright X--ray binaries.  For
background on X--ray binaries, see Chapter~1, and references therein.
For comprehensive reviews on black hole binaries, see Tanaka \& Lewin
(1995; hereafter TL95) and Tanaka and Shibazaki (1996; hereafter
TS96).  In this review, we emphasize the results of the past decade.
Throughout we make extensive use of the extraordinary data base
amassed since January 1996 by NASA's {\it Rossi X--ray Timing
Explorer} (Swank 1998).  Because of our page limit, several important
topics are omitted such as X--ray reflection studies (Done \&
Nayakshin 2001), the estimated number of black--hole X--ray novae in
the Galaxy (which is $\sim$1000; van den Heuvel 1992; TS96; Romani
1998), and the present and projected rates of discovery of black hole
binaries.

The main elements of the review are organized as follows.  In the
remainder of this section we catalog a total of 40 black--hole
binaries and candidate black--hole binaries and discuss several
introductory subjects, notably the physics of accretion onto black
holes.  In \S4.2 we present seven--year X-ray light curves for 20
systems.  In \S4.3 we provide modified definitions of the canonical
states of black hole binaries.  Both low-- and high--frequency
quasi--periodic oscillations are discussed in \S4.4.  A recurring
theme of the review is the importance of these black holes as
potential sites for tests of general relativity.

\subsection{The 18 black hole binaries}

The first black hole binary (BHB), Cygnus X--1, was established by
Webster and Murdin (1972) and Bolton (1972); the second, LMC X--3, was
identified by Cowley et al. (1983).  Both of these systems are
persistently bright in X--rays; furthermore, they are classified as
high--mass X--ray binaries (HMXBs) because their secondaries are
massive O/B stars (White et al. 1995).  The third BHB to be
established, A0620--00, is markedly different (McClintock \& Remillard
1986).  A0620--00 was discovered as an X--ray nova in 1975 when it
suddenly brightened to an intensity of 50 Crab\footnote{
1~Crab~=~$2.6~\times~10^{-9}$~erg~cm$^{-2}$~s$^{-1}$~keV$^{-1}$
(averaged over 2-11 keV)~ =~1.06~mJy~@~5.2~keV for a Crab--like
spectrum with photon index $\Gamma~=~2.1$.} to become the brightest
nonsolar X--ray source ever observed (Elvis et al. 1975).  Then, over
the course of a year, the source decayed back into quiescence to
become a feeble (1~{\it $\mu$}Crab) source (McClintock et al. 1995).
Similarly, the optical counterpart faded from outburst maximum by
$\Delta~V~\approx~7.1$ mags to $V~\approx~18.3$ in quiescence, thereby
revealing the optical spectrum of a K--dwarf secondary.

During the past 20 years, black holes (BHs) have been established in
15 additional X--ray binaries.  Remarkably, nearly all these systems
are X--ray novae like A0620--00.  Thus, in all there are now 18
confirmed BHBs. They are listed in Table~\ref{tab:bhb1} in order of
right ascension.  The coordinate name of each source is followed in
the second column by its variable star name (or other name, such as
Cyg~X--1), which is useful for web--based literature searches.  For
X--ray novae, the third column gives the year of discovery and the
number of outbursts that have been observed.  As indicated in the
table, among the confirmed BHBs, there are six recurrent X--ray novae
and three persistent sources, Cyg~X--1, LMC~X--3 and LMC~X--1.  As
indicated in the fourth column, these latter three sources are
high--mass X--ray binaries and are the only truly persistent sources
among the BHBs.

Two of the X--ray novae are peculiar: GRS~1915+105 has remained bright
for more than a decade since its eruption in August 1992; GX339--4
(1659--487), which was discovered by Markert et al. in 1973, undergoes
frequent outbursts followed by very faint states, but it has never
been observed to reach the {\it quiescent} state (Hynes et al. 2003).
Columns 5 and 6 give the peak X--ray flux and a distance estimate for
each source.  The orbital period is given in the seventh column (a
colon denotes an uncertain value) and the spectral type in the eighth.
The first one or two references listed for each binary contain
information on the orbital period and the spectral type.  The
remaining references support the distance estimates.

\begin{table*}
 \begin{center}
  \caption{Confirmed black hole binaries: primary properties}
   \label{tab:bhb1}
    \begin{tabular*}{\textwidth}{@{}l\x l\x l\x c\x r\x c\x r\x c\x c@{}}
   \hline \hline
Source    & Alternative& Year$^b$ &Type$^c$& F$_{\rm x,max}$&  D
           &P$_{\rm orb}$& 
           Spec.& References \\   
           &   name$^a$  &      &    &      ($\mu$Jy$^d$) &(kpc)& (hr)&
           & \\ \hline \hline

0422+32     &V518 Per  &1992/1  &L,T   & 3000   &$2.6\pm0.7$ &5.1
&M2V     &1,2\\
0538--641    &LMC X--3   &  --     &H,P   &   60   &$50\pm2.3$  &40.9
&B3V     &3,4\\
0540--697    &LMC X--1   &  --     &H,P   &   30   &$50\pm2.3$  &101.5
&O7III   &3,5,6\\
0620--003    &V616 Mon  &1975/2  &L,T   &50000   &$1.2\pm0.1$  &7.8
&K4V     &7,8,9,10\\
1009--45     &MM Vel    &1993/1  &L,T   &  800   &$5.0\pm1.3$  &6.8
&K7/M0V  &11,12\\

\vspace*{6 pt}
1118+480    &KV UMa    &2000/1  &L,T   &   40   &$1.8\pm0.5$ &4.1
&K5/M0V  &13,14\\
1124--684$^e$ &GU Mus    &1991/1  &L,T   & 3000   &$5\pm1.3$   &10.4
&K3/K5V  &15,15a,16\\
1543--475    &IL Lupi   &1971/4  &L,T   &15000   &$7.5\pm0.5$ &26.8
&A2V     &17,18\\
1550--564    &V381 Nor  &1998/5  &L,T   & 7000   &$5.3\pm2.3$ &37.0
&G8/K8IV &19\\    
1655--40     &V1033 Sco &1994/2  &L,T   & 3900   &$3.2\pm0.2$ &62.9
&F3/F5IV &20,21,22\\
1659--487$^f$ &V821 Ara &1972/$^f$  &L,T   & 1100   &4      &42.1:
&  --    &23,24\\

\vspace*{6 pt}
1705--250    &V2107 Oph &1977/1  &L,T   & 3600   &$8\pm2$     &12.5
&K3/7V   &7,25,26\\
1819.3--2525 &V4641 Sgr &1999/1  &L,T   &13000   &7.4--12.3  &67.6
&B9III   &27\\
1859+226    &V406 Vul  &1999/1  &L,T   & 1500   &11          &9.2:
&  --    &28,29\\
1915+105    &V1487 Aql &1992/1  &L,T   & 3700   &11--12       &804.0
&K/MIII  &30,31,32,33\\
1956+350    &Cyg X--1   &  --   &H,P   & 2300  &$2.0\pm0.1$ &134.4
&O9.7Iab &34,35\\
2000+251    &QZ Vul    &1988/1  &L,T   &11000   &$2.7\pm0.7$ &8.3
&K3/K7V  &35a,7,36,37\\
2023+338    &V404 Cyg  &1989/3  &L,T   &20000   &2.2--3.7     &155.3
&K0III   &38,39,40\\
   \hline \hline
  \end{tabular*}
 \end{center}
\begin{tabnote}
\tabnoteentry{a}{Name recognized by the SIMBAD Database and the
Astrophysics Data System (ADS).}
\tabnoteentry{b}{Year of discovery/number of outbursts observed (Chen
et al. 1997; this work).}
\tabnoteentry{c}{`H' -- HMXB, `L' -- LMXB, `T' -- transient, `P' --
persistent; Liu et al. 2000, 2001; this work.}
\tabnoteentry{d}{1~$\mu$Jy~=~$10^{-29}$~ergs~cm$^{-2}$~s$^{-1}$Hz$^{-1}$~=
$ 2.42~\times~10^{-12}$ergs~cm$^{-2}$~s$^{-1}$keV$^{-1}$.}
\tabnoteentry{e}{Commonly known as Nova Muscae 1991.}
\tabnoteentry{f}{Commonly known as GX339--4; number of outbursts
  $\sim10$ (Kong et al. 2002).} 
\tabnoteentry{}{\hskip-2.5ptREFERENCES:
$^{1}$Esin et al. 1997;
$^{2}$Filippenko et al. 1995a;
$^{3}$Freedman et al. 2001;
$^{4}$Cowley et al. 1983;
$^{5}$Hutchings et al. 1987;
$^{6}$Cowley et al. 1995;
$^{7}$Barret et al. 1996b;
$^{8}$Gelino et al. 2001b;
$^{9}$Marsh et al. 1994;
$^{10}$McClintock \& Remillard 2000;
$^{11}$Barret et al. 2000;
$^{12}$Filippenko et al. 1999;
$^{13}$McClintock et al. 2001a;
$^{14}$Wagner et al. 2001;
$^{15}$Orosz et al. 1996;
$^{15a}$Gelino et al. 2001a;
$^{16}$Shahbaz et al. 1997;
$^{17}$Orosz et al. 2002b;
$^{18}$Orosz et al. 1998;
$^{19}$Orosz et al. 2002a;
$^{20}$Hjellming \& Rupen 1995;
$^{21}$Orosz \& Bailyn 1997;
$^{22}$Shahbaz et al. 1999;
$^{23}$Cowley et al. 1987;
$^{24}$Hynes et al. 2003;
$^{25}$Remillard et al. 1996;
$^{26}$Filippenko et al. 1997;
$^{27}$Orosz et al. 2001;
$^{28}$Zurita et al. 2002;
$^{29}$Filippenko \& Chornock 2001;
$^{30}$Mirabel \& Rodriguez 1994;
$^{31}$Fender et al. 1999b;
$^{32}$Greiner et al. 2001a;
$^{33}$Greiner et al. 2001b;
$^{34}$Mirabel \& Rodrigues 2003;
$^{35}$Gies \& Bolton 1982;
$^{35a}$Harlaftis et al. 1996;
$^{36}$Filippenko et al. 1995b;
$^{37}$Casares et al. 1995;
$^{38}$Shahbaz et al. 1994;
$^{39}$Casares \& Charles 1994;
$^{40}$Casares et al. 1993.}

\end{tabnote}
\vspace*{-0.1in}
\end{table*}

The data in Table~\ref{tab:bhb1} reveal considerable diversity among
the BHBs.  For example, these 18 binaries range in size from tiny
XTE~J1118+480 with $P_{\rm orb} = 0.17$~days and a separation between
the BH and its companion of $a \approx 2.8$~\rsun~to GRS 1915+105 with
$P_{\rm orb} = 33.5$~days and $a \approx 95$\rsun.  Only six of these
18 systems were established as BHBs a decade ago (van Paradijs \&
McClintock 1995).  As indicated in Table~\ref{tab:bhb1}, all of the 15
X--ray novae are low--mass X--ray binaries (LMXBs), which typically
contain a secondary with a mass of roughly 1~\msun~or less (White et
al. 1995).  The BHBs 4U1543--47 and SAX~J1819.3--2525 have relatively
massive secondaries: $2.7 \pm 1.0$~\msun~and $2.9 \pm 0.2$~\msun,
respectively (Orosz et al. 2002b).  We classify them as LMXBs because
their secondary masses are comparable to the mass of the secondary of
Her~X--1 ($2.3~\pm~0.3$~\msun; Reynolds et al. 1997), which is a
well--known LMXB (Liu et al. 2001).  Furthermore, a
2--3~\msun~secondary is much less massive than the O/B secondaries
($\gtrsim$10~\msun) found in HMXB systems.

BHBs manifest themselves in five rather distinct spectral/temporal
states defined in the 1--10~keV band (e.g., van der Klis 1994; TL95;
TS96).  The three most familiar are (1) the {\it high/soft} (HS)
state, a high--intensity state dominated by thermal emission from an
accretion disk; (2) the {\it low/hard} (LH) state, a low--intensity
state dominated by power--law emission and rapid variability; and (3)
the {\it quiescent} state an extraordinarily faint state also
dominated by power--law emission.  The remaining two states, (4) the
{\it very high} (VH) state and (5) the {\it intermediate} state,
are more complex; recently they have come to the fore, as they have
now been observed in an appreciable number of sources.  These five BH
states and the transitions between them are the focus of \S4.3, where
they are redefined and illustrated in detail.

Additional data specific to the BH primaries are contained in
Table~\ref{tab:bhb2}. Of special importance is the mass function,
$f(M)~\equiv~P_{\rm orb}K_{2}^{3}/2\pi G~=~M_{1}$sin$^3i/(1+q)^{2}$.
The observables on the left side of the equation are the orbital
period, $P_{\rm orb}$, and the half--amplitude of the velocity curve
of the secondary, $K_{2}$.  On the right, the quantity of most
interest is the BH mass, $M_{1}$; the other parameters are
the orbital inclination angle, $i$, and the mass ratio,
$q~\equiv~M_{2}/M_{1}$, where $M_{2}$ is the mass of the secondary.
Thus a secure value of the mass function can be determined for a
quiescent X--ray nova or an HMXB by simply measuring
the radial velocity curve of the secondary star.  The mass function
values are given in the second column of Table~\ref{tab:bhb2}.
An inspection of the equation for $f(M)$ shows that the value of the
mass function is the absolute minimum mass of the compact primary.
Thus, for 12 of the 18 BHBs, the very secure value of f(M) alone is
sufficient to show that the mass of the compact X--ray source is at
least 3~\msun~(Table~\ref{tab:bhb2}), which is widely agreed to exceed
the maximum stable mass of a neutron star (NS) in GR (Rhoades \&
Ruffini 1974; Kalogera \& Baym 1996). For the remaining half--dozen
systems, some additional data are required to make the case for a
BH. The evidence for BHs in these 18 systems is generally very strong
(see Ch. 5).  Thus, assuming that GR is valid in the strong--field
limit, we choose to refer to these compact primaries as BHs, rather
than as BH candidates.  We note, however, our reservations about three
of the systems listed in Table~\ref{tab:bhb2}: (1) The mass function
of LMC~X--1 (0540-697) is quite uncertain; (2)
the orbital period of XTE~J1859+226 is not firmly established
(Filippenko \& Chornock 2001; Zurita et al. 2002); and (3) the
dynamical data for GX~339--4 (1659-487) were determined in outburst by
a novel technique, and neither the orbital period nor the velocity
amplitude are securely determined (Hynes et al. 2003).

\begin{table*}
\begin{center}
  \caption{Confirmed black hole binaries: X--ray and optical data}
  \label{tab:bhb2}
  \begin{tabular*}{\textwidth}{@{}l\x c\x c\x c\x c\x c\x c\x l\x l@{}}
   \hline \hline
Source      &f(M)$^a$      & M$_{1}^a$  &f(HFQPO)  &f(LFQPO) &Radio$^b$&
E$_{max}^c$ &References \\
        &(M$_{\odot}$)&(M$_{\odot}$) &(Hz)     &(Hz)
&&(MeV)& \\ \hline \hline

0422+32    &1.19$\pm$0.02 &3.2--13.2&    --     & 0.035--32  & P & 0.8,1--2:& 
1,2,3,4,5\\

0538--641   &2.3$\pm$0.3   &5.9--9.2&    --      &   0.46     & --  & 0.05&
6,7\\

0540--697   &0.14$\pm0.05$  &4.0--10.0:&    --   & 0.075      & --   & 0.02& 
8,7\\

0620--003   &2.72$\pm$0.06 &3.3--12.9&    --     &    --       & P,J? & 0.03:&
9,10,11,11a\\

1009--45    &3.17$\pm$0.12 &6.3--8.0&    --    & 0.04--0.3 & --$^d$&0.40, 1:& 
12,4,13\\

\vspace*{6 pt}
1118+480   &6.1$\pm$0.3   &6.5--7.2&    --      & 0.07--0.15 &   P  & 0.15& 
14,15,16,17\\

1124--684   &3.01$\pm$0.15 &6.5--8.2&    --      & 3.0-8.4    & P      & 0.50& 
18,19,20,21\\

1543--475   &0.25$\pm$0.01 &7.4--11.4$^e$&    -- &    7       & -- $^f$& 0.20&
22,4\\

1550--564   &6.86$\pm$0.71 &8.4--10.8&92,184,276& 0.1-10     & P,J    & 0.20& 
23,24,25,26,27\\

1655--40    &2.73$\pm$0.09 &6.0--6.6&300,450    & 0.1--28    & P,J    & 0.80& 
28,29,30,31,54\\

1659--487   &$>$~2.0$^{g}$  & --  &    --      & 0.09--7.4  & P      & 0.45, 1:& 
32,33,4,13\\

\vspace*{6 pt}
1705--250   &4.86$\pm$0.13 &5.6--8.3&    --      &     --      & --$^d$ & 0.1&
34,35\\

1819.3--2525&3.13$\pm$0.13 &6.8--7.4&    --      &     --      & P,J    & 0.02&
36,37\\

1859+226   &7.4$\pm$1.1   &7.6--12:&    190   & 0.5--10    & P,J?   & 0.2& 
38,39,40,41\\

1915+105   &9.5$\pm$3.0   &10.0--18.0:&  41,67,113,168 &   0.001-10 & P,J&0.5, 1:& 
42,43,44,4,13\\

1956+350   &0.244$\pm$0.005 &6.9--13.2&    --   & 0.035--12  & P,J& 2--5& 
45,46,47,48,49\\

2000+251   &5.01$\pm$0.12 &7.1--7.8&    --      & 2.4--2.6   & P & 0.3&
18,50,51\\

2023+338   &6.08$\pm$0.06 &10.1--13.4&   --    &      --      & P & 0.4&
52,53\\
   \hline \hline
  \end{tabular*}
 \end{center}
\begin{tabnote}
\tabnoteentry{a}{Orosz et al. 2002b, except for 1659--487; colon denotes
uncertain value.}
\tabnoteentry{b}{Radio properties: `P' - persistent over 10 or more
days and/or inverted spectrum; `J' - relativistic jet detected.}
\tabnoteentry{c}{Maximum energy reported; colon denotes uncertain value.}
\tabnoteentry{d}{No observations made.}
\tabnoteentry{e}{Orosz, private communication.}
\tabnoteentry{f}{Very faint (e.g., see IAUC 7925).}
\tabnoteentry{g}{For preferred period, $P~=~1.76$~days, 
$f(M)~=~5.8~\pm~0.5$~\msun; Hynes et al. 2003.}
\tabnoteentry{}{\hskip-2.5ptREFERENCES:
$^1$van der Hooft et al. 1999;
$^2$Vikhlinin et al. 1992; 
$^3$Shrader et al. 1994;
$^4$Grove et al. 1998;
$^5$van Dijk et al. 1995;
$^6$Boyd et al. 2000;
$^7$Nowak et al. 2001;
$^8$Ebisawa et al. 1989;
$^9$Owen et al. 1976;
$^{10}$Kuulkers et al. 1999;
$^{11}$Coe et al. 1976;
$^{11a}$ Marsh et al. 1994;
$^{12}$van der Hooft et al. 1996;
$^{13}$Ling et al. 2000;
$^{14}$Wood et al. 2000;
$^{15}$Revnivtsev et al. 2000b; 
$^{16}$Fender et al. 2001;
$^{17}$McClintock et al. 2001a;
$^{18}$Rutledge et al. 1999;
$^{19}$Belloni et al. 1997;
$^{20}$Ball et al. 1995;
$^{21}$Sunyaev et al. 1992;
$^{22}$This work;
$^{23}$Remillard et al. 2002b;
$^{24}$Corbel et al. 2001;
$^{25}$Wu et al. 2002;
$^{26}$Corbel et al. 2003;
$^{27}$Sobczak et al. 2000b;
$^{28}$Remillard et al. 1999;
$^{29}$Strohmayer 2001a;
$^{30}$Hjellming \& Rupen 1995;
$^{31}$Hannikainen et al. 2000;
$^{32}$Revnivtsev et al. 2001; 
$^{33}$Corbel et al. 2000;
$^{34}$Wilson \& Rothschild 1983;
$^{35}$Cooke et al. 1984;
$^{36}$Hjellming et al. 2000;
$^{37}$Wijnands \& van der Klis 2000;
$^{38}$Cui et al. 2000a;
$^{39}$Markwardt 2001;
$^{40}$Brocksopp et al. 2002;
$^{41}$Dal Fiume et al. 1999;
$^{42}$Morgan et al. 1997;
$^{43}$Strohmayer 2001b;
$^{44}$Mirabel \& Rodriguez 1994;
$^{45}$Vikhlinin et al. 1994;
$^{46}$Cui et al. 1997b;
$^{47}$Stirling et al. 2001;
$^{48}$Ling et al. 1987;
$^{49}$McConnell et al. 2002;
$^{50}$Hjellming et al. 1988;
$^{51}$Sunyaev et al. 1988;
$^{52}$Han \& Hjellming 1992;
$^{53}$Sunyaev et al. 1991b;
$^{54}$Tomsick et al. 1999.}

\end{tabnote}
\vspace*{-0.1in}
\end{table*}

The mass of a BH can be derived from the measurement of its mass
function in combination with estimates for $M_2$ and sin~$i$.  The
ratio of the mass of the BH to the mass of the secondary star is
usually deduced by measuring the rotational velocity of the latter.
The binary inclination angle can be constrained in several ways;
commonly, one models the photometric variations associated with the
gravitationally-distorted secondary star that is seen to rotate once
per binary orbit in the plane of the sky (e.g., Greene, Bailyn, \&
Orosz 2001). Mass estimates for the known BHBs are given in the third
column of Table~\ref{tab:bhb2}.  In an astrophysical environment, a BH
is completely specified in general relativity by two numbers, its mass
and its specific angular momentum or spin, $a = J / c M_1$, where $J$
is the BH angular momentum and $c$ is the speed of light (e.g., Kato
et al. 1998). The spin value is conveniently expressed in terms of a
dimensionless spin parameter, $a_{*} = a / r_g$, where the
gravitational radius is $r_g \equiv GM/c^{2}$. The value of $a_{*}$
lies between 0 for a Schwarzschild hole and 1 for a
maximally--rotating Kerr hole. The radius of the BH event horizon
depends on both $M_1$ and $a_{*}$; this topic is discussed further in
\S4.1.5.

High frequency QPOs (HFQPOs) in the range 40--450~Hz are observed for
the four systems with data given in the fourth column.  Low--frequency
QPOs (LFQPOs) are also observed from these systems and ten others, as
indicated in column~5.  Most of the systems were at one time or another
detected as radio sources and at least five have exhibited resolved
radio jets (column~6; see Mirabel \& Rodriguez 1999). An X--ray jet
(two--sided) has been observed from XTE~J1550--564 (Corbel et
al. 2002).  As shown in column 7, power--law emission extending to
$\sim$1~MeV has been observed for six sources.  The references listed
in the last column support the X--ray QPO, radio, and $E_{max}$ data
given in columns 4--7.

The dynamical evidence for massive ($M~>~3$~\msun) collapsed stellar 
remnants is indisputable.  However, this evidence alone can never
establish the existence of BHs.  At present, the argument for BHs
depends on the assumption that GR is the correct theory of strong
gravity.  To make an airtight case that a compact object is a BH with
an event horizon, we must make clean quantitative measurements of
relativistic effects that occur near the collapsed
object.  That is, we must measure phenomena that are predicted by GR
to be unique to BHs.  A possible route to this summit is to measure
and interpret the effects of strong--field gravity that are believed
to be imprinted on both HFQPOs (\S4.4.3) and X--ray emission lines
(\S4.2.3).  Another approach, which is discussed in \S4.3.4, is based
on sensing the qualitative differences between an event horizon and
the material surface of a NS.

\subsection{Black hole candidates}

Certain characteristic X--ray properties of the 18 established BHs are
often used to identify a candidate BH when the radial velocities of
the secondary cannot be measured (e.g., because the secondary is too
faint).  These frequently observed characteristics of the established
BHBs, which are discussed in detail by TL95, include an ultrasoft
X--ray spectrum (1--10~keV), a power--law spectral tail that extends
beyond 20 keV, characteristic state transitions (\S4.1.2; \S4.3), and
rapid temporal variability.  However, none of these putative BH
signatures has proved to be entirely reliable; each of them has been
forged by one or more systems known to contain a NS primary (TL95).
This is not surprising since the X--ray spectral/temporal properties
originate in an accretion flow that is expected to be fairly similar
whether the primary is a BH or a weakly magnetized NS.  Another
characteristic of all BHBs is a complete absence of either periodic
pulsations or type I X--ray bursts, which are the very common
signatures of NS systems.

\begin{table*}
\begin{center}
  \caption{Candidate black hole binaries$^a$}
  \label{tab:bhc}
  \begin{tabular*}{\textwidth}{@{}l\x l\x l\x c\x c\x c\x r@{}}
   \hline \hline
Source&            RA(2000)~~~~~&    DEC(2000)& $r_{\rm x}$$^b$& 
BH trait$^c$& Grade$^d$&  References \\ \hline

1354--645 (BW Cir)    & 13 58 09.74  &-64 44 05.2  &           &LH,HS
& A & 1,2,3,4\\
1524--617 (KY TrA)    & 15 28 16.7   &-61 52 58    &           &LH,HS 
& A & 5,6,7 \\
4U~1630--47           & 16 34 01.61  &-47 23 34.8  &           &LH,HS
& A & 8,9,10,11,83 \\
XTE J1650--500        & 16 50 01.0   &-49 57 45    &           &LH,HS,VH 
& A & 12,13,14,15,16 \\
SAX J1711.6--3808     & 17 11 37.1   &-38 07 06    &           &LH,HS 
& B & 17,18 \\
\vspace*{6 pt}
GRS 1716--249$^e$         & 17 19 36.93  &-25 01 03.4  &           &LH 
& B & 19,20,21 \\
XTE J1720--318        & 17 19 59.06  & -31 44 59.7 &           &LH:,HS 
& C & 22,23,24 \\
KS 1730--312          & 17 33 37.6   &-31 13 12    &$30''$     &LH,HS 
& C & 25,26 \\
GRS 1737--31          & 17 40 09     &-31 02.4     &$30''$     &LH 
& B & 27,28,29 \\
GRS 1739--278         & 17 42 40.03  &-27 44 52.7  &           &LH,HS,VH 
& A & 30,31,32,33,34 \\
1E 1740.7--2942       & 17 43 54.88  &-29 44 42.5  &           &LH,HS,J
& A & 35,36,37,38,39 \\
\vspace*{6 pt}
H 1743--322          & 17 46 15.61   &-32 14 00.6  &           &HS,VH 
& A & 40,41,42,80,81,82 \\
A 1742--289           & 17 45 37.3   &-29 01 05    &           &HS:
& C & 43,44,45,46 \\
SLX 1746--331         & 17 49 50.6   &-33 11 55    &$35''$     &HS: 
& C & 47,48,49 \\
XTE J1748--288        & 17 48 05.06  &-28 28 25.8  &           &LH,HS,VH,J
& A & 50,51,52,53,54\\
XTE J1755--324        & 17 55 28.6   &-32 28 39    &$1'$       &LH,HS 
& B & 55,56,57,58\\
1755--338 (V4134 Sgr) & 17 58 40.0   &-33 48 27    &           &HS
& B & 59,42,60,61,62 \\
\vspace*{6 pt}
GRS 1758--258         & 18 01 12.67  &-25 44 26.7  &           &LH,HS,J
& A & 63,38,64,65,66 \\
EXO 1846--031         & 18 49 16.9   &-03 03 53    &$11''$$^f$ &HS
& C & 67 \\
XTE J1908+094        & 19 08 53.08  &+09 23 04.9  &           &LH,HS
& B & 68,69,70,71 \\
1957+115 (V1408 Aql)~& 19 59 24.0   &+11 42 30    &           &HS
& C & 72,42,73,74,75 \\
XTE J2012+381        & 20 12 37.70  &+38 11 01.2  &           &LH,HS 
& B & 76,77,78,79 \\ 
   \hline \hline
  \end{tabular*}
 \end{center}
\begin{tabnote}
\tabnoteentry{a}{For additional references and information, see TL95
and Liu et al. 2001.}
\tabnoteentry{b}{Positional uncertainty given if $r_{\rm x}~>~10''$.}
\tabnoteentry{c}{`LH' -- {\it low/hard} state, `HS' -- {\it high/soft}
state, `VH' -- {\it very high} state, `J' -- radio jet.}
\tabnoteentry{d}{Qualitative grade indicating the likelihood that the
candidate is in fact a BH.}
\tabnoteentry{e}{GRS~1716--249~=~GRO~J1719--24~=~X--ray Nova Oph 1993.}
\tabnoteentry{f}{Alternative position also possible; see Parmar et al.
1993.}
\tabnoteentry{}{\hskip-2.5ptREFERENCES: 
$^{1}$Kitamoto et al. 1990; $^{2}$Brocksopp et al. 2001; 
$^{3}$Revnivtsev et al. 2000a; $^{4}$McClintock et al. 2003b;
$^{5}$Murdin et al. 1977; $^{6}$Kaluzienski et al. 1975;
$^{7}$Barret et al. 1992; $^{8}$Hjellming et al. 1999;
$^{9}$Tomsick \& Kaaret 2000; $^{10}$Dieters et al. 2000;
$^{11}$Augusteijn et al. 2001; $^{12}$Groot et al. 2001;
$^{13}$Kalemci et al. 2002; $^{14}$Homan et al. 2003a; 
$^{15}$Miller et al. 2002b; $^{16}$Sanchez--Fernandez et al. 2002;
$^{17}$in 't Zand et al. 2002a; $^{18}$Wijnands \& Miller 2002;
$^{19}$Mirabel et al. 1993; $^{20}$van der Hooft et al. 1996;
$^{21}$Revnivtsev et al. 1998b; $^{22}$Rupen et al. 2003;
$^{23}$Remillard et al. 2003a; $^{24}$Markwardt \& Swank 2003;
$^{25}$Borozdin et al. 1995; $^{26}$Vargas et al. 1996;
$^{27}$Ueda et al. 1997; $^{28}$Cui et al. 1997a; 
$^{29}$Trudolyubov et al. 1999; $^{30}$Marti et al. 1997;
$^{31}$Vargas et al. 1997; $^{32}$Borozdin et al. 1998;
$^{33}$Wijnands et al. 2001; $^{34}$Greiner et al. 1996;
$^{35}$Cui et al. 2001; $^{36}$Churazov et al. 1993;
$^{37}$Smith et al. 1997; $^{38}$Smith et al. 2002;
$^{39}$Marti et al. 2000; $^{40}$Gursky et al. 1978;
$^{41}$Cooke et al. 1984; $^{42}$White \& Marshall 1984;
$^{43}$Davies et al. 1976; $^{44}$Wilson et al. 1977;
$^{45}$Branduardi et al. 1976; $^{46}$Kennea \& Skinner 1996;
$^{47}$Skinner et al. 1990; $^{48}$White \& van Paradijs 1996;
$^{49}$Motch et al. 1998; $^{50}$Hjellming et al. 1998b;
$^{51}$Revnivtsev et al. 2000c; $^{52}$Kotani et al. 2000;
$^{53}$Miller et al. 2001; $^{54}$Rupen et al.1998;;
$^{55}$Remillard et al. 1997; $^{56}$Ogley et al. 1997;
$^{57}$Revnivtsev et al. 1998a; $^{58}$Goldoni et al. 1999;
$^{59}$Bradt \& McClintock 1983; $^{60}$White et al. 1988;
$^{61}$Pan et al. 1995; $^{62}$Seon et al. 1995;
$^{63}$Rodriguez et al. 1992; $^{64}$Sunyaev et al. 1991a;
$^{65}$Smith et al. 2001; $^{66}$Rothstein et al. 2002;
$^{67}$Parmar et al. 1993; $^{68}$Rupen et al. 2002;
$^{69}$in't Zand et al. 2002b; $^{70}$Woods et al. 2002;
$^{71}$Chaty et al. 2002; $^{72}$Margon et al. 1978;
$^{73}$Wijnands et al. 2002;
$^{74}$Yaqoob et al. 1993; $^{75}$Nowak \& Wilms 1999;
$^{76}$Hjellimg et al. 1998a; $^{77}$Campana et al. 2002;
$^{78}$Vasiliev et al. 2000; $^{79}$Hynes et al. 1999;
$^{80}$Revnivtsev et al. 2003; $^{81}$Steeghs et al. 2003;
$^{82}$Homan et al. 2003b;
$^{83}$Remillard \& McClintock 1999.}

\end{tabnote}
\end{table*}

Nevertheless, despite the limitations of these spectral/temporal
identifiers, they have served as useful guides and have allowed the
identification of a number of probable BH primaries, which we refer to
as BH candidates or BHCs.  In order to economize on acronyms, we also
use BHC to refer to the binary system that hosts the BH candidate.  A
list of 22 BHCs is given in Table~\ref{tab:bhc}. These systems are
less well--known than the BHBs listed in Table~\ref{tab:bhb1}.
Therefore, to aid in their identification we have included in most
cases the prefix to the coordinate source name that identifies the
discovery mission (e.g., EXO~=~{\it EXOSAT}).  For the four sources
with variable star names, the prefix is omitted.  In columns 2--4, we
give the best available coordinates and their uncertainties (if they
exceed $10''$).  These coordinates are also an excellent resource for
interrogating ADS and SIMBAD.

In column~5 we list the characteristics that indicate the BHC
classification, e.g., an observation of one or more of the canonical
states of a BHB.  A source classified as ``ultrasoft'' in the older
literature (e.g., White \& Marshall 1984) is assumed here to have been
observed in the HS state.  Three BHCs with resolved radio jets are
also noted in column~5.  Variability characteristics (\S4.4) and broad
Fe K$\alpha$ emission lines (\S4.2.3) can also be important BH
indicators, but these are not explicitly noted in
Table~\ref{tab:bhc}. Selected references are cited in the last column.
The coordinate data can be found in the first reference.  The
following few references support the BH characteristics given in
column~5. For several sources, a few additional references provide
limited information on optical/IR/radio counterparts, etc.

We have assigned a grade to each BHC that is based both on how
thoroughly the candidate has been observed and on the characteristics
it displayed.  In fact, this grade is qualitative and
subjective and meant only as a guide.  Our sense of the grade is as
follows: We would be surprised if even one of the nine A--grade BHCs
contains a NS, but not surprised if one of the six C--grade BHCs
contains a NS.  In compiling Table~\ref{tab:bhc}, we have been
selective.  For example, we did not include SAX~J1805.5--2031 (Lowes
et al. 2002) and XTE~J1856+053 (Barret et al. 1996a) primarily because
we judged that the available information was too scanty.  Thus the
total number of BH or BHC cases considered here is 40.  For narrative
discussions about many of the systems in Tables~\ref{tab:bhb1} and
\ref{tab:bhc}, see TL95.  For additional data and as a
supplement to the list of references given in Table~\ref{tab:bhc}, see
Liu et al. (2001).

\subsection{X--ray novae}

If we include the X--ray novae observed during the past decade, then
about 300 bright binary X--ray sources are known (van Paradijs 1995;
Liu et al. 2000; Liu et al. 2001). More than half of them are LMXBs,
and roughly half of each type (i.e., LMXB and HMXB) are classified as
transient sources.  The HMXB transients are neutron--star/Be--star
binaries, which are not relevant to this review.  The well--studied
LMXB transients, on the other hand, include all of the X--ray novae
that are listed in Tables~\ref{tab:bhb1}--\ref{tab:bhc}.  The
principal hallmarks of an X--ray nova include both the discovery of
the source during a violent outburst and a very large ratio of maximum
to minimum X--ray intensity.  The behavior of A0620--00, described in
\S4.1.2 provides a classic example of an X--ray nova.  Indeed, it is
the extreme faintness of the quiescent accretion disk in these systems
that allows one to view the companion star, leading to secure
dynamical measurements of the BH mass (\S4.1.2, \S4.3.4; Ch. 5).

Recurrent eruptions have been observed for several of the X--ray novae
listed in Table~\ref{tab:bhb1}: e.g., A0620--00 in 1917 and 1975;
H1743--322 in 1977 and 2003; GS~2023+338 in 1938, 1956 and 1987;
4U~1543--47 in 1971, 1983, 1992 and 2002; and 4U~1630--472 and
GX~339--4 at $\sim1-2$ year intervals.  (Outbursts prior to 1970 are
inferred from studies of optical plates ex post facto.)  Most X--ray
novae, however, have been observed to erupt only once.  Nevertheless,
all X--ray novae are thought to be recurrent, with cycle times for
some possibly as long as several centuries or more. TS96 suggest an
average cycle time of 10--50 years.  The infrequent outbursts are due
to a sudden surge in the mass accretion rate onto the BH.  The
generally accepted mechanism driving the outburst cycle is that
described by the disk instability model, which was developed initially
for dwarf novae (Smak 1971; Osaki 1974; Cannizzo 1993; Lasota et
al. 2001) and extended to X--ray novae (e.g., Dubus et al. 2001).

\subsection{Accretion onto black holes}

The desire to understand observations of BHBs compels us to model the
hydrodynamics and radiation processes of gas orbiting in the
gravitational potential of a compact object (see Ch. 13 for a detailed
review).  The best--known such model is the thin accretion disk
(Pringle \& Rees 1972; Shakura \& Sunyaev 1973; Novikov \& Thorne
1973; Lynden--Bell \& Pringle 1974). For nearly all of the systems
included in Tables~\ref{tab:bhb1}--\ref{tab:bhc}, the companion star
fills its Roche equipotential lobe and a narrow stream of gas escapes
the star through the inner Lagrangian ($L_{1}$) point.  This gas has
high specific angular momentum and cannot accrete directly onto the
BH.  It feeds into a thin disk of matter around the BH known as an
accretion disk.  Once entrained in the disk, the gas moves in
Keplerian orbits with angular velocity $(GM/R^{3})^{1/2}$.  However,
viscous dissipation slowly taps energy from the bulk orbital motion,
and viscosity transports angular momentum outward.  As a result, the
gas gets hotter as it sinks deeper into the gravitational potential
well of the BH.  Near the BH the disk terminates because there are no
stable particle orbits possible in the extreme gravitational field.
The existence of an innermost stable circular orbit (ISCO) and other
properties of BHs are discussed in many texts (e.g.  Shapiro \&
Teukolsky 1983; Kato et al. 1998).  A defining property of a BH is its
event horizon, the immaterial surface that bounds the interior region
of spacetime that cannot communicate with the external universe. The
radius of the event horizon of a Schwarzschild BH ($a_{*} = 0$) is
\RS~$\equiv$~2\Rg~$\equiv~2(GM/c^{2})~=~30$~km($M/$10\msun), the ISCO
lies at $R_{\rm ISCO}$~=~6\Rg, and the corresponding maximum orbital
frequency is $\nu_{\rm ISCO}~=~220$~Hz($M/$10\msun)$^{-1}$ (see
\S4.1.2 for the definition of $a_{*}$).  For an extreme Kerr BH
($a_{*} = 1$), the radii of both the event horizon and the minimum
stable (prograde) orbit are identical, $R_{\rm K}~=~R_{\rm
ISCO}$~=~\Rg, and the maximum orbital frequency is $\nu_{\rm
ISCO}~=~1615$~Hz($M/$10\msun)$^{-1}$.  For the Kerr BH, it is well
known that the rotational energy can be tapped electromagnetically
(Blandford \& Znajek 1977).  The gas flows driven by this process are
both anisotropic and self--collimating (Blandford 2002, and references
therein), and they may be the source of the relativistic jets seen
from several BHBs and BHCs (Tables~\ref{tab:bhb2}--\ref{tab:bhc}).

Even for XTE~J1118+480, the smallest system in Table~\ref{tab:bhb1},
the outer radius of the accretion disk is expected to be roughly one
solar radius, $\sim10^{5}$\Rg, vastly larger than the BH event
horizon.  Thus a gas element of mass $m$ that is destined to enter the
BH starts far out with negligible binding energy.  However, when it
reaches the ISCO it will have radiated $0.057mc^{2}$ for a
Schwarzschild BH or $0.42mc^{2}$ for an extreme Kerr BH.  Moreover,
90\% of this colossal binding energy is radiated within about 20\Rg~of
the center.  At all disk radii, the binding energy liberated by
viscous dissipation is radiated locally and promptly and results in a
gas temperature that increases radially inward reaching a maximum of
$T\sim~10^{7}$~K near the BH. This picture is the basis of the
standard thin accretion disk model (Shakura \& Sunyaev 1973).  A
nonrelativistic approximation to this thin disk spectrum has been
formulated conveniently as a multi--temperature blackbody (Mitsuda et
al. 1984; Makishima et al. 1986). The total disk luminosity in a
steady state is $L_{\rm disk}~=~GM$\mdot/$2R_{\rm in}$, where \mdot~is
the mass accretion rate and $R_{\rm in}$ is the radius of the inner
edge of the disk.  This model, often referred to as the
Multicolor Disk (MCD) model, is used to describe the thermal component
that is dominant in the HS state and is also present in the VH state
(\S4.2.2; \S4.3.5; \S4.3.7).  The MCD model has been available within
XSPEC (``diskbb''; Arnaud \& Dorman 2002) for many years and has been
widely used.  Despite the successes of the MCD model, it is important
to note a significant limitation, namely, the neglect of a
torque--free boundary condition at the ISCO (Gierlinski et al. 1999).
In the MCD model, the temperature profile is simply $T(R) \propto
R^{-3/4}$, which rises to a maximum at the ISCO, whereas a proper
inner boundary condition produces null dissipation at the ISCO with
the temperature peaking at $R > R_{\rm ISCO}$ (Pringle 1981;
Gierlinski et al. 1999).  The current MCD model requires attention and
improvement.

Efforts have also been made to include relativistic corrections to the
MCD model (e.g., Ebisawa et al. 1991; Zhang et al. 1997a; Gierlinski
et al. 2001).  However, quantitative analyses (e.g., a spectroscopic
measurement of the inner disk radius when the source distance and
inclination angle is known) will require more sophisticated
models. Accretion disk models are being developed to incorporate MHD
effects in the context of GR (e.g., McKinney \& Gammie 2002), with
additional considerations for radiation pressure and radiative
transfer (Turner et al. 2002; Shimura \& Takahara 1995).  Early
results show that magnetic fields may couple matter in the ``plunging
region'' to matter at radii greater than the ISCO, and thereby extract
energy from very near the horizon (Agol \& Krolik 2000).  In the case
of a rapidly--rotating Kerr hole, the spin/electromagnetic effects
mentioned above may not only drive relativistic jets, but also may
modify grossly the spectrum of the inner accretion disk (Wilms et
al. 2001b; Miller et al. 2002b).

At lower mass accretion rates corresponding to several percent of the
Eddington luminosity, a BHB usually enters the LH (i.e., {\it
low/hard}) state and at very low accretion rates it reaches the {\it
quiescent} state, which may be just an extreme example of the LH
state.  In both of these states, the spectrum of a BHB is dominated by
a hard, nonthermal power--law component (photon index $\sim$1.7;
\S4.3.4 \& \S4.3.6), which cannot be accounted for by a thermal
accretion disk model; it is most plausibly explained as due to
Comptonization of soft photons by a hot optically--thin plasma.  Early
models for the spectrum of the {\it quiescent} state postulated that
the disk does not extend all the way down to the ISCO (Narayan 1996;
Narayan et al. 1996).  The disk is truncated at some larger radius and
the interior volume is filled with a hot ($T_{e} \sim 100$~keV)
advection--dominated accretion flow or ADAF (Narayan \& Yi 1994, 1995;
Narayan et al. 1996; Quataert \& Narayan 1999).  In an ADAF, most of
the energy released via viscous dissipation remains in the accreting
gas rather than being radiated away (as in a thin disk).  The bulk of
the energy is advected with the flow.  Only a small fraction of the
energy is radiated by the optically thin gas before the gas reaches
the center. Consequently, the radiative efficiency of an ADAF (which
depends on the uncertain fraction of the viscous energy that is
channeled to the electrons) is expected to be only $\sim$0.1--1\%,
whereas the radiative efficiency discussed above for disk accretion is
definitely $\geq~5.7$\%.  There is wide agreement ñthat these
radiatively inefficient flows have been observed in quiescent BHBs
(\S4.3.4) and from galactic nuclei (Baganoff et al. 2001; Loewenstein
et al. 2001).  However, the theoretical picture has become complex
with variant models involving winds (ADIOS; Blandford \& Begelman
1999) and convection (CDAFs; Igumenshchev \& Abramowicz 1999; Stone et
al. 1999; Narayan et al. 2000; Quataert \& Gruzinov 2000)

\begin{figure}
\epsfxsize2.5in 
\epsfysize2.5in 
\hspace*{-0.20in}
\hspace*{+0.08in}
\epsfbox{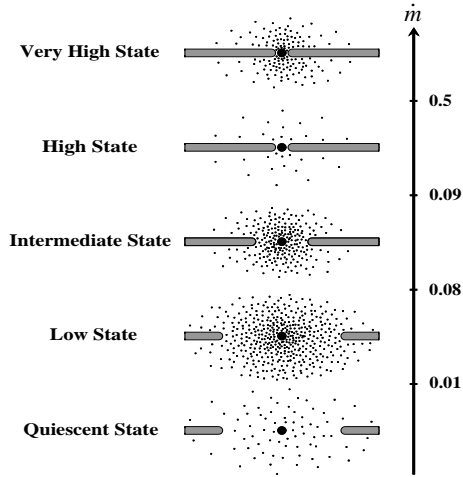}
\caption{Schematic sketch of the accretion flow in different spectral
states as a function of the total Eddington--scaled mass accretion rate
\eddmdot.  The ADAF is represented by dots and the thin disk by the
horizontal bars.  The {\it very high} state is illustrated, but it is
not included in the unification scheme (Esin et al. 1997).}
\label{adaf_esin}
\vspace*{-0.07in}
\end{figure}

One attempt has been made to unify four of the five states of a BHB
using both the MCD and ADAF models (Esin et al. 1997).  This approach
is illustrated in Figure~\ref{adaf_esin}, which shows how the geometry
of the accretion flow changes as the mass accretion rate
\eddmdot~varies (\eddmdot~is the mass accretion rate expressed in
Eddington units).  The scenario indicates how a BH system progresses
through five distinct states of increasing \eddmdot~from the {\it
quiescent} state to the VH state.  In the three states at lower
\eddmdot, the flow consists of two zones (disk and ADAF), as described
above.  For the two states of highest \eddmdot, the disk extends down
to the ISCO.  In all five states, the disk is bathed in a corona that
is a seamless continuation of the ADAF.  Apart from the VH state, the
model treats consistently the dynamics of the accreting gas, the
thermal balance of the ions and electrons in the ADAF and corona, and
the radiation processes.  The model has had significant successes in
describing the spectral evolution of several BHBs (Esin et al. 1997;
Esin et al. 1998; for further discussion of the ADAF model, see
\S4.3.4.)

However, the multistate model of Esin et al. (1997) has important
limitations.  For example, it does not unify the most luminous state,
the {\it very high} state, which is characterized by an unbroken
power--law spectrum extending out to a few hundred keV or more
(\S4.3.7).  Also, this simple ordering of the states by
\eddmdot~or luminosity is naive (\S4.3).  Moreover, the model does not
account for the dynamic behavior of the corona, including strong
flares and powerful low--frequency quasi--periodic oscillations
(\S4.4), nor does it account for the radio emission observed from most
BHBs (Table~\ref{tab:bhb2}).  Finally, the ``evaporation'' process
by which the cold gas in a thin disk feeds into a hot ADAF is at best
qualitatively understood, and there is no quantitative model relating
the disk truncation radius to the accretion rate \eddmdot~ (Narayan
2002, and references therein).

There are alternative models of the X--ray states, and many of them
invoke a dynamic accretion disk corona that is fed by MHD
instabilities in the disk.  For example, in the model of Merloni \&
Fabian (2001a, 2001b) the hot corona that generates the power--law
component is intimately connected with the thin accretion disk.
Magnetic energy generated (presumably) by the sheared Keplerian disk
creates magnetic flares that rise out of the disk because of the
Parker instability.  Within the framework of this model, Di Matteo et
al. (1999) present a magnetic flare model for the two common states of
GX339--4.  In the HS state, the flares occur near the disk and heat
it.  The disk reradiates the observed soft thermal component, whereas
the faint hard component is produced by Comptonization of the soft
flux.  In the LH state, the flares occur far above the disk and the
density of soft seed photons is greatly reduced.  Thus, the system is
photon--starved, and the resultant Comptonized spectrum is hard.
Merloni and Fabian (2002) also consider coronae as sources of powerful
jets/outflows.  They find that such outflows can render a source
radiatively inefficient even if advection of energy into the BH is
unimportant.

It is generally agreed that the temperatures in the corona are in the
range 100--300~keV, with optical depths of 0.1--1 (e.g., Merloni \&
Fabian 2001b).  There is little agreement, however, on the geometry
and physical properties of the corona.  Thus, a wide range of coronal
models have been proposed (e.g., Haardt \& Maraschi 1991; Dove et
al. 1997a, 1997b; Meyer et al. 2000; Rozanska \& Czerny 2000;
Kawaguchi et al. 2000; Nowak et al. 2002; Liu et al. 2002).  Liu et
al. conclude that the primary difficulty in modeling the corona is the
magnetic field that produces time variations and spatial
inhomogeneities; in addition, one must consider complicated
radiation/energy interactions between the disk and the corona.
Because of the complexity of the problem, some students of the corona
choose to apply their models to large quantities of X--ray (and radio)
data, an approach that has proved fruitful (Zdziarski et
al. 2002, 2003; Wardzinski et al. 2002).  

Timing studies have been used to isolate the fast X--ray variability
in BHBs as primarily emanating from the corona rather than from the
accretion disk (Churazov et al. 2001). Timing studies have also
created requirements to explain X--ray QPOs, such as the strong
low--frequency QPOs that are prevalent in the VH state. Many coronal
models do not deal with fast variability or QPOs explicitly, although
there are exceptions, such as the ``Accretion--Ejection Instability''
model (Tagger \& Pellat 1999).  In this model, magnetic spiral waves
heat and accelerate material at those locations where the spiral waves
and the disk are in corotation.  During BHB outbursts, it has become
exceedingly clear that spectral and timing characteristics at both
X--ray and radio frequencies may change dramatically and
abruptly. Therefore, no single model can account for all of the
complex behavior, and our goal in this review is to summarize what is
known about each X--ray state and to then discuss physical models in
the context of a given state.

In addition to the models discussed above, we mention briefly the jet
model (e.g., Falcke \& Biermann 1995). This model is motivated by the
observations of resolved radio and X--ray jets
(Tables~\ref{tab:bhb2}--\ref{tab:bhc}), by observations of
radio/X--ray correlations (e.g., Hannikainen et al. 2001; Corbel et
al. 2000), and by successes in modeling the broadband spectra of some
systems as synchrotron radiation.

\subsection{Some consequences of an event horizon}

The properties of BHs and BH accretion flows are discussed in many
texts (e.g., Shapiro \& Teukolsky 1983; Kato et al. 1998; Abramowicz
1998).  As mentioned above, a defining feature of a BH is its event
horizon.  Since BHs lack a material surface, some effects observed for
NSs (e.g., type I bursts) are absent for a BH.
Similarly, a BH cannot sustain a magnetic field anchored within it,
and hence they cannot generate periodic X--ray pulsations, which are
observed for many NSs.

Both type I bursts and periodic pulsations are considered firm
signatures of a NS (TL95).  It is interesting to ask what fraction of
cataloged, bright sources show either pulsations or type~I bursts.  We
examined this question using the catalog of van Paradijs (1995),
selecting only the brighter sources (F$_{\rm x}~>~30~\mu$Jy) that have
been optically identified.  We excluded the confirmed and candidate BH
systems listed in Tables~\ref{tab:bhb1}--\ref{tab:bhc}.  For the 21
HMXB systems that met our selection criteria, we found that 18 out of
21 (86\%) pulse.  For 21 LMXB systems, we found that 12 burst and
2 pulse (67\% burst or pulse).  Thus a very high fraction of these
sources manifest behavior that identifies them as NSs.  It is also
interesting to consider which systems have failed to produce
detectable bursts or pulsations because they are either an unusual NS
source or they contain a BH.  The three non--pulsing HMXB sources are
1700--37, 1947+30 and Cyg~X--3; the six corresponding LMXB sources
are LMC~X--2, 1543--62, Sco X--1, GX349+2, GX9+9 and 1822--00.

\subsection{The Rossi X--ray Timing Explorer (RXTE): 1996.0 -- present}

{\it RXTE} has been in continuous operation since its launch on 1995
December 30.  Its prime objective is to investigate the fundamental
properties of BHs, NSs and white dwarfs by making high time resolution
observations of extremely hot material located near BH event horizons
or stellar surfaces.  It is the largest X--ray detector array ever
flown, and it provides energy coverage from 2--200~keV.  It features
high throughput (up to $\sim$150,000 counts~s$^{-1}$) and $\sim1~\mu$s
time resolution.  {\it RXTE's} year--round, wide--sky coverage and its
fast response time has made it an important vehicle for the study of
transient phenomena and for the support of multiwavelength science.

The observatory is comprised of two large area instruments (PCA and
HEXTE) that act in concert, viewing the sky through a common $1^{\rm
o}$ field of view.  The third instrument is an All--Sky Monitor (ASM)
that surveys about 80\% of the sky each orbit.  For descriptions of
the instruments, see Levine et al. (1996), Swank (1998), Rothschild et
al. (1998), and Bradt et al. (2001).  The Proportional Counter Array
(PCA), which has a total net area of 6250~cm$^{2}$, is the chief
instrument.  The PCA consists of five sealed proportional--counter
detectors.  It is effective over the range 2--60~keV with 18\% energy
resolution at 6~keV.  The High Energy Timing Experiment (HEXTE) covers
the energy range 20--200~keV.  It is comprised of eight NaI/CsI
phoswich detectors with a combined net area of 1600~cm$^{2}$.  The
HEXTE has provided important spectral information beyond the reach of
the PCA, but the modest count rates of the HEXTE (e.g., 289
counts~s$^{-1}$ for the Crab) limit its use for timing studies.
Indeed, it is the PCA (12,800 counts~s$^{-1}$ for the Crab with 5
PCUs) that has achieved groundbreaking results in high--energy
astrophysics.

The ASM is comprised of three wide--field proportional counters that
are mounted on a rotating boom.  On a daily basis, it surveys about
90\% of the sky and obtains about 5--10 observations per source.  It
locates bright sources to a typical accuracy of $\sim$5$'$.  In
uncrowded regions it can monitor known sources down to about 35 mCrab
(2~$\sigma$) in one 90--s exposure or about 10 mCrab in one day.  On
many occasions, the ASM has proved indispensable in alerting PCA/HEXTE
observers to targets of opportunity, such as the appearance of a new
transient or a state--change in a cataloged source.  Furthermore, the
ASM public archive containing the continuous light curves of $\sim$400
X--ray sources (both Galactic and extragalactic) has been invaluable
in studying the multiyear behavior of X--ray sources and in
supplementing X--ray and multiwavelength (e.g., {\it Chandra} and {\it
HST}) studies of individual sources.  The data are also important to
observers of AGN and gamma--ray bursts.

\section{X--ray light curves, spectra and luminosity data}

\subsection{ASM light curves of BH binaries and BH candidates}

\begin{figure}
\epsfxsize29.55pc 
\epsfysize29.55pc 
\hspace*{-0.39in}
\hspace*{-0.23in}
\epsfbox{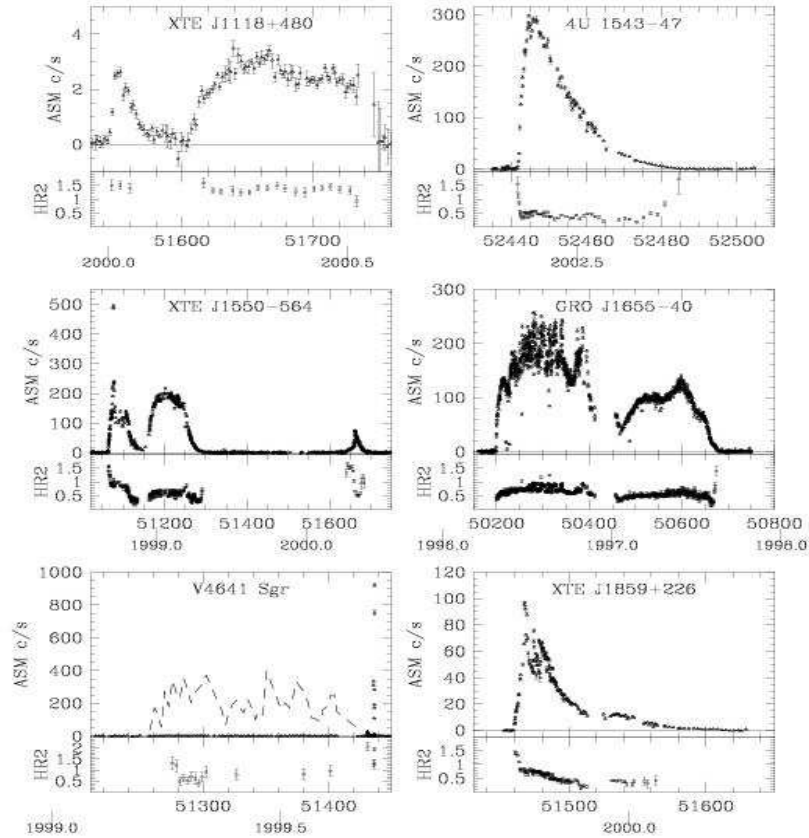}
\vspace*{-0.51in}
\caption{Transient ASM light curves of six black hole binaries.  For
V4641 Sgr, the dashed line shows intensity x50 in order to highlight the
low-level activity that preceded the violent and short--lived flare.}
\label{asm_bh}
\end{figure}

\begin{figure}
\epsfxsize29.55pc 
\epsfysize29.55pc 
\hspace*{-0.59in}
\hspace*{-0.16in}
\epsfbox{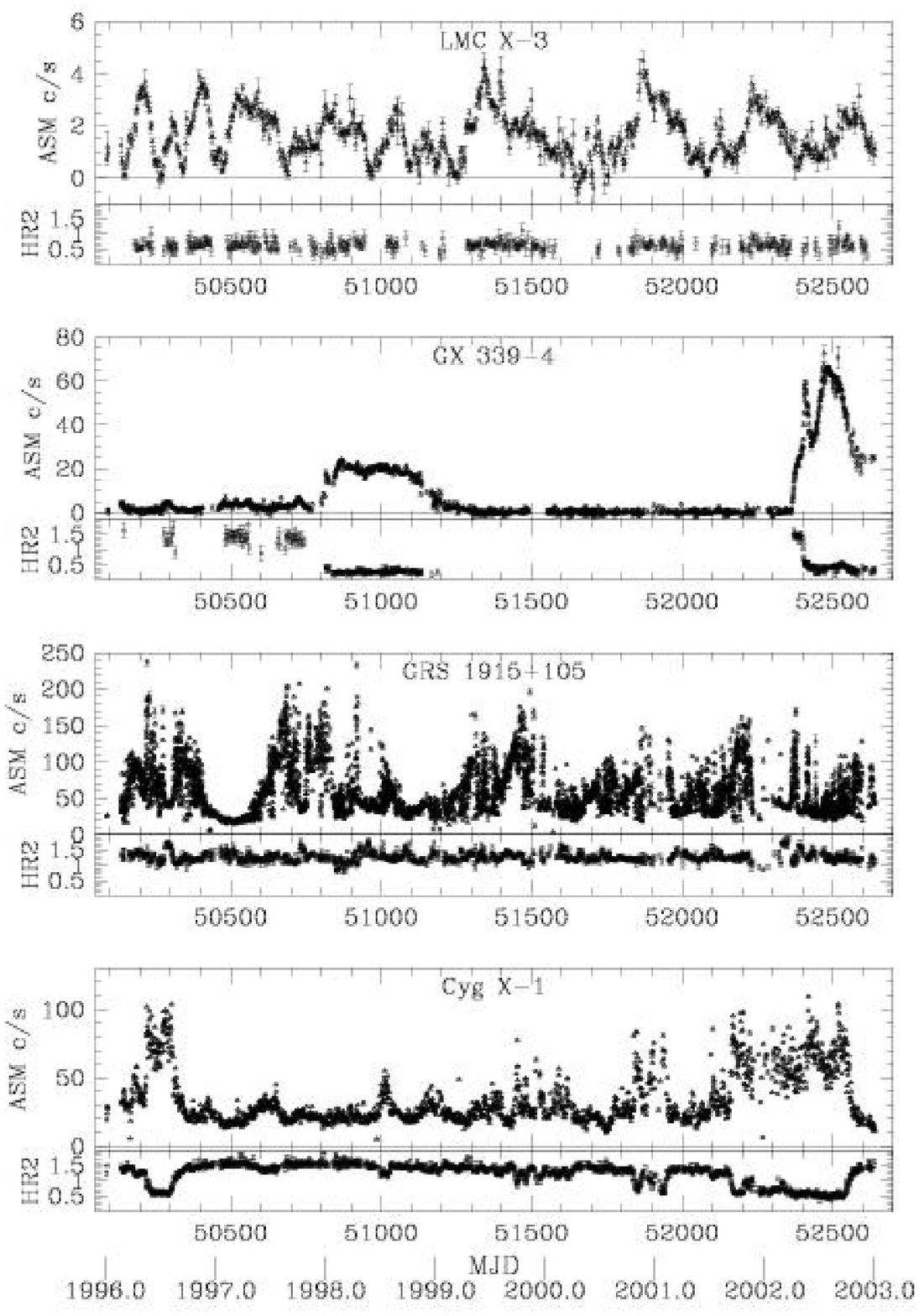}
\vspace*{-0.47in}
\caption{Seven--year ASM light curves of four black hole binaries.}
\label{asm_bh_pers}
\end{figure}

To illustrate the diversity of behavior among BHBs and BHCs, we show
in Figure~\ref{asm_bh}--\ref{asm_bhc_pers} the 1.5--12~keV ASM light
curves and the (5--12~keV)/(3--5~keV) hardness ratio (HR2) for 20 of
the systems listed in Tables~\ref{tab:bhb1}--\ref{tab:bhc} that were
active during the past seven years.  These light curves should be
compared to the heterogeneous collection of $\sim$20 X--ray light
curves collected by Chen et al. (1997).  In making such comparisons,
note that we use a linear intensity scale, whereas Chen et al. and
most authors use a log scale.  The light curves in
Figures~\ref{asm_bh}--\ref{asm_bhc_pers} are ordered by RA, although we
discuss them roughly in order of increasing complexity.  An intensity
of 1 Crab corresponds to 75.5~ASM c~s$^{-1}$.  A hardness ratio (HR2)
of 0.5 (1.5) generally corresponds to the HS state (LH state).  All
good data are included, although the time interval for binning the
counts has been tailored to the source intensity.  The hardness ratio
is not plotted in the absence of a detectable 5-12~keV
flux. References are cited sparingly in the narrative descriptions of
the light curves given below; see Tables~\ref{tab:bhb1}--\ref{tab:bhc}
for further references.  In the following, the term ``classic'' light
curve refers to an outburst profile that exhibits a fast rise and an
exponential decay, like those observed for several pre--{\it RXTE} BH
X--ray novae, including A0620--00, GS/GRS~1124--68, GS~2000+25 and
GRO~J0422+32 (TL95; Chen et al. 1997).  Such classic light curves often
show a secondary maximum (roughly a doubling of intensity) that occurs
during the decline phase about 40--100 days after the onset of the
outburst (TL95; Chen et al. 1997).

\subsubsection{Black hole binaries}

Figure~\ref{asm_bh} shows the light curves of all six of the X--ray
novae with short outburst cycles that were detected by the ASM
(Tables~\ref{tab:bhb1}--\ref{tab:bhb2}).  {\it 4U~1543--47:} An
exceptionally clean example of a classic light curve with an
e--folding decay time of $\approx14~$days. It lacks a secondary
maximum, presumably because the outburst is so brief.  For details and
references on three prior outbursts of this source, see Chen et
al. (1997).  {\it XTE~J1859+226:} A second example of a classic light
curve that does show a secondary maximum (at about 75 days after
discovery).  Note the intense variability near the primary maximum.
{\it XTE~J1118+480:} One of five X--ray novae that remained in a hard
state throughout the outburst and failed to reach the HS state. Note
the prominent precursor peak.

Also shown in Figure~\ref{asm_bh} are the complex and similar light
curves of two X--ray novae with long orbital periods, GRO~J1655--40
and XTE~J1550--564.  {\it GRO~J1655--40:} This source has undergone
two outbursts since its discovery in 1994 July.  Shown here is the
full light curve of the second, 16--month outburst.  The
double--peaked profile is quite unlike the classic profile of
4U~1543--47.  During the first maximum in 1996, the source exhibited
strong flaring and intense nonthermal emission (VH state).  In 1997
the source spectrum was soft and thermal except for a hard episode at
the very end of the outburst (Sobczak et al. 1999).  Note the several
absorption dips in the light curve of this high inclination system
(Kuulkers et al. 1998). {\it XTE~J1550--564:} The complex profile
includes two dominant peaks during 1998--99, followed several hundred
days later by a smaller peak in 2000.  Not shown here are three very
small outbursts (LH states) that have occurred subsequently.  Some
unusual characteristics include the slow 10--day rise following the
source's discovery on 1998 September 6, followed by the dominant
X--ray flare (6.8~Crab; September 19-20), and the abrupt $\sim$10--day 
decay timescale following each outburst. The source was predominately 
in the VH or {\it intermediate} states during the first peak (1998), 
softening to the HS state during
the second peak (1999). The spectral evolution through all of the
X--ray states occurred much more rapidly during the small peak in
2000.  {\it SAX~J1819.3--2525~=~V4641~Sgr:} As this extraordinary
light curve shows, the source became active at a low level in the
spring of 1999 (dashed line shows intensity x50).  Five months later
it underwent a brief, violent flare during which the 1.5--12~keV
intensity increased very rapidly (within 7 hours) from 1.6 to 12.2
Crab.  Within two hours thereafter, the intensity declined to less
than 0.05~Crab (Wijnands \& van der Klis 2000, and references
therein).

Figure~\ref{asm_bh_pers} shows the light curves of the two persistent
BHBs (LMC~X--3 and Cyg~X--1) and two other BHBs that have been active
throughout the {\it RXTE} era.  {\it LMC~X--3:} The light curve shows
the large--amplitude cyclic modulation in the flux reported by Cowley
et al.  (1991); however, their $\sim$198--day cycle time is shorter
than is indicated by these data.  As the hardness ratio plot shows,
the source remains in the HS state most of the time.  However, at the
local intensity minima, where there are gaps in the HR2 plot due to
statistical limitations, the PCA observations show transitions to the
LH state (Wilms et al. 2001a).  {\it GX~339--4:} As shown, the source
underwent major eruptions into a soft spectral state
(HR~$\approx~0.5$) in 1998 and 2002.  In the time between these two
outbursts, the source was very faint ($<$~2~mCrab) compared to its
customary hard--state intensity of $\sim$30 mCrab, which it enjoyed
prior to its 1998 outburst.  Remarkably, this transient BHB has never
been observed in a fully {\it quiescent} state (Hynes et al. 2003).
{\it GRS~1915+105:} This source exhibits extraordinary variations in
both the X--ray and radio bands.  Astonishing and yet repetitive
patterns are sometimes seen in the X--ray light curves (Muno et
al. 1999; Belloni et al. 2000; Klein--Wolt et al. 2002).  For a 1992
X--ray light curve showing the birth of this source, see Chen et
al. (1997).  {\it Cyg~X--1:} Transitions between the LH and HS states
were first observed in this archetypal BH source (\S4.3.1).  However,
as this record shows, there are both gradual and rapid variations in
the hardness ratio that suggest both rapid state transitions and
intermediate conditions between the HS and hard states (see \S4.3.9).

\subsubsection{Black hole candidates}

\begin{figure}
\epsfxsize29.55 pc
\epsfysize29.55 pc
\hspace*{-0.43in}
\hspace*{-0.22in} 
\epsfbox{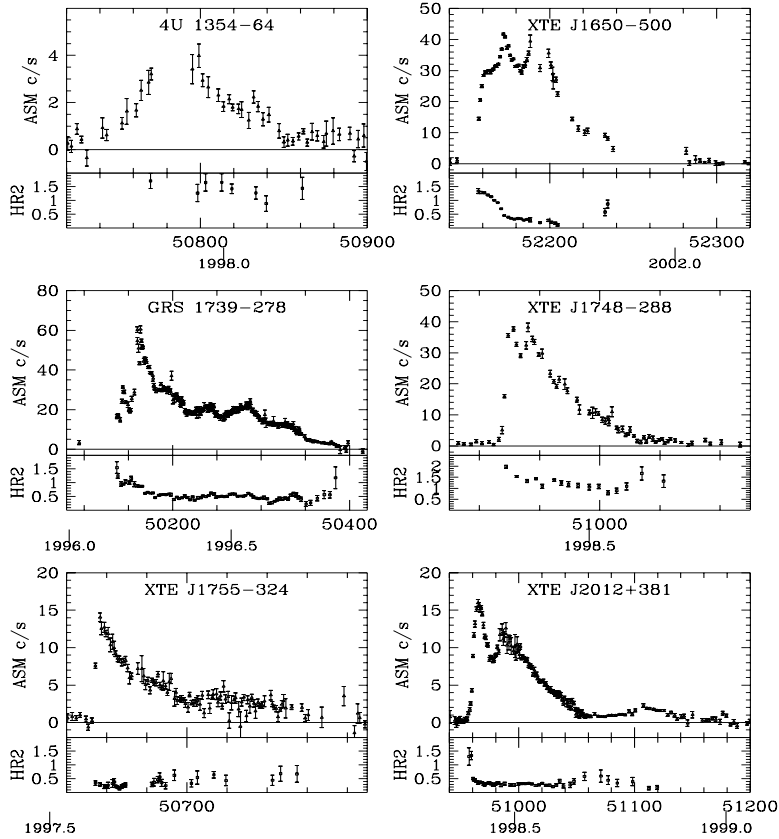}
\vspace*{-0.55in}
\caption{Transient ASM light curves of six black hole candidates.}
\label{asm_bhc}
\end{figure}

Figure~\ref{asm_bhc} displays the light curves of six BHCs with short
outburst cycles.  {\it XTE J748--288:} This short duration outburst
that begins and ends in a hard spectral state is similar to the
classic light curve of 4U~1543--47 (Fig.~\ref{asm_bh}). The 1/e decay
time is $\approx16$~days.  This source is heavily absorbed, and the
values of the hardness ratio are consequently increased.  {\it
GRS~1739--278:} A somewhat longer duration outburst with a nearly
classic profile that includes a precursor peak, $\sim$40\%
variability near maximum, and undulations in intensity during the
decay.  Again, the outburst begins and ends with a hard
spectrum.  {\it XTE~J1755--324:} This brief outburst, which follows an
abrupt rise, provides yet another example of a classic light
curve. This outburst is locked in the HS state. {\it
XTE~J2012+381:} This unusual light curve combines a classic rise to
maximum followed directly by a precipitous drop in intensity.  Also
unusual are a large secondary maximum occurring just 30~days after
discovery plus an additional late maximum at $\sim$140~days.  {\it
XTE~J1650--500:} A complex light curve.  At the onset of the outburst
there is a very rapid rise followed by a slow rise.  This unusual
behavior is accompanied by a remarkable, slow ($\sim$15--day)
transition from a hard spectral state to a soft one.  {\it
4U~1354--64:} A slow rise followed by a rather rapid decline.  During
this outburst, the source remained in the LH state (Brocksopp et
al. 2001), whereas the HS state was reached during a brighter outburst
in 1987 (Kitamoto et al. 1990).  This source may be identical to Cen
X--2, which reached a peak intensity of 13 Crab in 1967 (Brocksopp et
al. 2001).

\begin{figure}
\epsfxsize29.55 pc
\epsfysize29.55 pc
\hspace*{-0.63in}
\hspace*{-0.17in}
\epsfbox{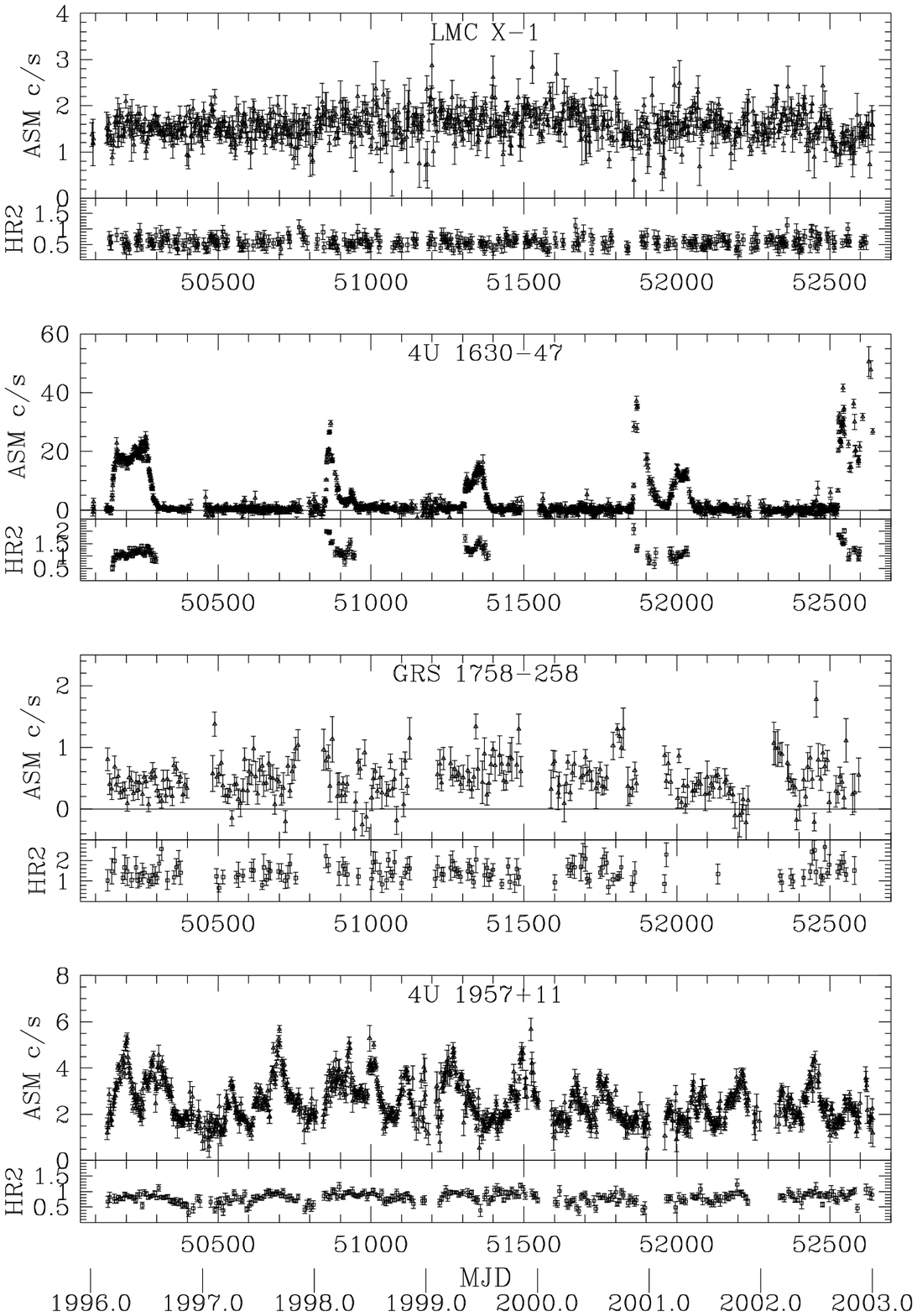}
\vspace*{-0.39in}
\caption{Seven--year ASM light curves of three black hole candidates
and the black hole binary LMC~X--1.}
\label{asm_bhc_pers}
\end{figure}

In Figure~\ref{asm_bhc_pers} we show the light curves of the BHB
LMC~X--1 and the light curves of three BHCs. {\it LMC~X--1:} The
source is continually in a soft spectral state and maintains a
relatively steady intensity.  {\it 4U~1630--47:} The $\sim$600--day
recurrence time has been known for 25 years (Jones et al. 1976).  Here
we see five, nearly equally--spaced outbursts.  Note the very
different profiles and fluences of the outbursts.  Note also that the
1996 outburst starts with a soft spectrum.  {\it
GRS1758--258:} This hard Galactic center source was discovered by
GRANAT/SIGMA in 1991.  Its spectrum extends to at least 300~keV
(Sunyaev et al. 1991a).  During 2001 it underwent a transition to an
unusual soft state of very low intensity (Smith et al. 2001; Miller et
al. 2002d); the low flux level during that event is evident in the ASM
record shown here.  {\it 4U~1957+11:} The source has long been
considered a BHC based on its ``ultrasoft'' spectrum (White \&
Marshall 1984), although Yaqoob et al. (1993) have argued that the
primary is a NS. The source displays a consistent flaring behavior and
a soft spectrum over the 7--year interval.

It is often said that during its initial rise the spectrum of a BH
X--ray nova transitions from a hard spectral state to a soft one.  For
several sources, the data in Figures~\ref{asm_bh}--\ref{asm_bhc_pers}
support this view: 4U~1543--47, XTE~J1550--564, XTE~J1859+226,
XTE~J1650--500, GRS~1739--278, and XTE~J2012+381.  However, there are
two clear counter--examples, sources whose spectra {\it hardened}
during the initial rise: GRO~J1655--40 (Fig.~\ref{asm_bh}) and
4U~1630--47 (Fig.~\ref{asm_bhc_pers}; first of 5 outbursts).

\subsection{Synoptic studies of selected black hole binaries}

It is important to follow the several--month spectral evolution of
individual sources and to construct unified spectral models that can
be used to represent the energy spectra of all BHBs.  The necessary
elements of such models can be deduced from a simple appraisal of the
observational data: e.g., Cyg X--1 and other BHBs in the LH state show
that the model must contain a nonthermal component, which can be well
represented by a power--law function (TL95).  On the other hand, the
soft spectra observed for most BH X--ray novae in the HS state is most
widely modeled as a multi--temperature blackbody, which approximates
the emission from an optically--thick (relativistic) accretion
disk. Many studies of spectral evolution therefore choose a
composite model comprised of disk blackbody and power--law
components. Although this simple model has significant limitations,
nevertheless it has proven to be widely applicable and quite effective
in monitoring the spectra of BHBs, as we now discuss briefly by
pointing to two examples.

Important studies of {\it Ginga} spectra using this model were made by
Ebisawa et al. (1991, 1993, 1994).  The authors added one 
refinement to the model, namely, a broad absorption feature above
7~keV.  This feature is associated with the reflection of X--rays by
an optically thick accretion disk (Ebisawa et al. 1994). 
With this model a successful and quantitative
comparison was made of the spectra of Cyg~X--1, LMC~X--3, GS~2000+25,
LMC~X--1, GX339--4 and GS/GRS~1124--68 (Nova Mus 1991).  This latter
source, which exhibited a classic light curve, was observed 51 times
in 1991 over a span of 235 days using the {\it Ginga} Large Area
Counter (LAC).  We direct the reader's attention to Figure 15 in
Ebisawa et al. (1994) which shows the evolution over a full outburst
cycle of the spectral parameters and fluxes for GS/GRS~1124--68.  In
their figure, it is evident that an important
transition occurs 130 days into the outburst. For
example, the photon spectral index suddenly decreases from 2.2--2.6 to
a value near 1.6, the hard flux increases substantially and the soft
disk flux decreases.  This characteristic behavior, which has been
observed for a number of BH X--ray novae, marks the transition from
the HS ({\it high/soft}) state to the LH ({\it low/hard}) state.

We now compare the results obtained for GS/GRS~1124--68 to the results
of an analogous study of the irregular BH X--ray nova XTE~J1550--564,
which was observed extensively by {\it RXTE} during its 1997--1998
outburst (Sobczak et al. 2000b).  A total of 209 pointed observations
spanning the entire 255--day outburst were made using the PCA and
HEXTE detectors. The 1998--1999 light curve of the source is complex
and includes a slow (10~day) rise to maximum, an intense (6.8~Crab)
flare that occurred early in the outburst, and a ``double--peaked''
profile that roughly separates the outburst into two halves of
comparable intensity (Fig. 4.2).  Sobczak et al. adopted very nearly
the same spectral model and methodology as Ebisawa et al. (1994).  We
direct the reader's attention to Figure~4 and the accompanying text in
Sobczak et al. (2000b), which describes a complex course of evolution
relative to that of GS/GRS~1124--68.  Very briefly, Sobczak et
al. show that during the first half of the outburst QPOs are
ubiquitous and the spectrum is dominated by the power--law component,
which are conditions that mark the {\it very high/intermediate} state.
In contrast, during the second half of the outburst QPOs are scarce
and the spectrum is dominated by emission from the accretion disk,
which corresponds to the {\it high/soft} state.  During this state,
the inner disk radius (\S4.1.5) remained nearly constant for 4 months
(Sobczak et al. 2000b).  Very similar behavior has been observed for
GS/GRS~1124--68 (Ebisawa et al. 1994) and for several other sources
(TL95).  The constancy of the disk inner radius is remarkable, given
the accompanying, large variations in luminosity that are usually
observed (TL95).

In overview, in both GS/GRS~1124--68 and XTE~J1550--564, one sees time
intervals where thermal emission from the disk dominates the spectrum,
while at other times the disk spectrum is substantially modified and a
{\it power--law component may dominate at either high or low
luminosity}. In other sources, such as Cyg X--1, the nonthermal LH
state may be stable for many months or years, but its soft--state
spectrum does not resemble the thermal spectra seen from
GS/GRS~1124--68 and XTE~J1550--564. These results highlight the need
to specify the physical properties of X--ray states while avoiding the
confusing terminology that has developed during the past 30 years.
Accordingly, in \S4.3 we address this issue and seek clearer
definitions of the X--ray states.

\subsection{Relativistic iron emission lines}

Strong evidence for accretion disks in active galactic nuclei has come
from X--ray observations of broad iron K$\alpha$ lines.  In
particular, in some Seyfert galaxies the very asymmetric profile of
the Fe K$\alpha$ line (e.g., its extended red wing) suggests strongly
that the emission arises in the innermost region of a relativistic
accretion disk (for reviews see Fabian et al. 2000, Reynolds \& Nowak
2003).  The good energy resolution of {\it ASCA} provided the first
clear evidence for such a line profile (Tanaka et al. 1995).  In some
cases, the line profile indicates the presence of an accretion disk
extending down to the ISCO (Weaver et al. 2001).  The broad Fe
K$\alpha$ fluorescence line is thought to be generated through the
irradiation of the cold (weakly--ionized) disk by a source of hard
X--rays (likely an optically--thin, Comptonizing corona).
Relativistic beaming and gravitational redshifts in the inner disk
region can serve to create an asymmetric line profile.

In fact, the first broad Fe K$\alpha$ line observed for either a BHB
or an AGN was reported in the spectrum of Cyg X--1 based on {\it
EXOSAT} data (Barr et al. 1985).  It was this result that inspired
Fabian et al. (1989) to investigate the production of such a line in
the near vicinity of a Schwarzschild BH, a result that was later
generalized by Laor (1991) to include the Kerr metric.  Other early
studies of relativistically smeared Fe K$\alpha$ lines from BHBs were
conducted by Done et al. (1992) and others.  Their work was one part
of a broader examination of the accretion geometry that is produced as
hard X--rays from an overlying corona illuminate an optically--thick
accretion disk.  An Fe K$\alpha$ line and a reflected continuum are
always generated in this case (George \& Fabian 1991; Matt et
al. 1991).  One example is a {\it Ginga} study of the reflected
spectrum of V404~Cyg (Zycki et al. 1999a, 1999b).  A limitation of
this study is the use of proportional counter detectors (e.g., the
{\it Ginga} LAC and {\it RXTE} PCA, which have an energy resolution of
only FWHM~$\approx~1.2$~keV at Fe K$\alpha$).  Such iron K$\alpha$
studies suffer both because the energy resolution is marginal and
because the response matrices of the detectors are uncertain at the
1--2 \% level, while the Fe line profile in BHBs is typically only
1--5 \% above the X--ray continuum. Consequently, the results from
these instruments must be considered with caution. Some iron K$\alpha$
sources that have been studied with {\it RXTE} include: GRO~J1655--40
(Balucinska--Church \& Church 2000); XTE~J1748--288 (Miller et
al. 2001); and GX~339--4 (Feng et al. 2001; Nowak et al. 2002).

\begin{figure}
\epsfxsize20pc 
\epsfysize20pc 
\hspace*{-0.35in}
\hspace*{0.12in}
\epsfbox{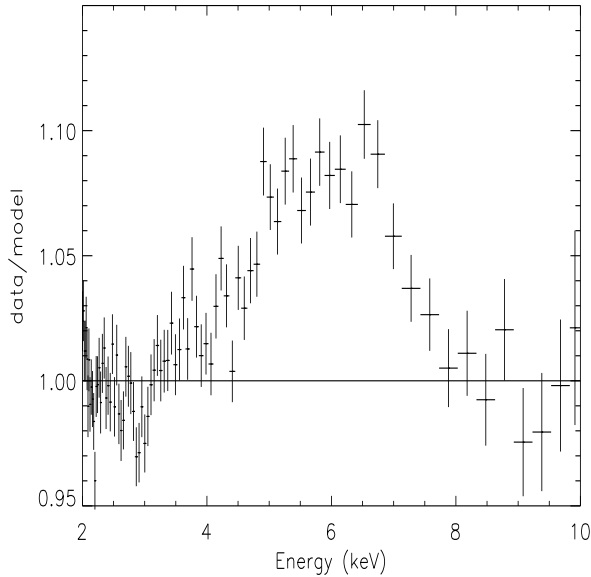}
\vspace*{-0.12in}
\caption{Data/model ratio for XTE~J1650--500.  The model consists of
multicolor disk blackbody and power--law components (Miller et
al. 2002b).  Note the non--Gaussian shape 
and low--energy extent of the line profile.}
\label{miller}
\vspace*{-0.07in}
\end{figure}

More telling studies of the Fe K$\alpha$ line have been achieved using
the MECS and HP--GSPC detectors aboard {\it BeppoSAX}, which have a
resolution of $\approx~0.6~$keV at Fe~K$\alpha$.  Broad line profiles
($\sim$4--9~keV) have been observed for SAX J1711.6--3808 (in't Zand
et al. 2002a) and XTE J1908+094 (in't Zand et al.  2002b); however,
these profiles are rather symmetric and may be more a product of
Compton scattering than relativistic broadening.  {\it BeppoSAX}
studies have also revealed other systems with broad, asymmetric Fe~K
profiles that resemble the predictions of relativistic smearing:
GRS~1915+105 (Martocchia et al. 2002) and V4641~Sgr (Miller et
al. 2002a).  Interestingly, for both of these systems the inner disk
radius deduced from the line profile is consistent with the radius of
the ISCO for a Schwarzschild BH, suggesting that rapid spin is not
required.

Spectral studies at higher resolution have been made recently with
{\it XMM--Newton} and {\it Chandra}.  Using the {\it XMM} EPIC--MOS1
detector, Miller et al. (2002b) find for XTE~J1650--500 a broad,
skewed Fe K$\alpha$ emission line (Fig.~\ref{miller}) which suggests
the presence of an extreme Kerr BH and indicates a steep radial
falloff of disk emissivity with radius.  An observation of Cyg X--1
with the HETGS grating and ACIS--S detector aboard {\it Chandra}
revealed a broad line centered at $\approx~5.82$~keV with a FWHM of
$\approx~1.9$~keV (Miller et al. 2002c).  Also present was a smeared
Fe edge at $\approx~7.3$~keV.  The authors conclude that the line is
predominately shaped by Doppler/gravitational effects and to a lesser
degree by Compton scattering due to reflection.

\subsection{Super--Eddington luminosities}

Recently there has been considerable interest in ultraluminous X--ray
sources (ULXs) in external galaxies with 0.5--10~keV luminosities in
the range $10^{39}-10^{40.5}$ erg~s$^{-1}$ (Makishima et al. 2000;
Fabbiano et al. 2001; Humphrey et al. 2003; Miller et al. 2003b;
Ch. 12).  The luminosities of ULXs greatly exceed the Eddington limit
of a 1.4~\msun~NS: $L_{\rm Edd}~=
1.25~\times~10^{38}(M_{1}/M_{\odot})$~\ergsec. The most luminous
systems also exceed by a factor of $\sim$20 the Eddington luminosity
of a typical 10\msun~BH.  This fact has led to the suggestion that the
most luminous ULXs are a new class of accreting BHs with masses
$\sim$100~\msun~(Makishima et al. 2000; Fabbiano et al. 2001).
Alternatively, it has been suggested that the ULXs are powered by
conventional stellar--mass BHs that radiate anisotropically (King et
al. 2001; Ch. 13).  We examine this question by comparing as directly
as possible the luminosities of the ULXs to the luminosities of the 18
BHBs listed in Tables~\ref{tab:bhb1}--\ref{tab:bhb2}.  We also mention
briefly apparent differences between ULXs and BHBs in their
spectra and duty cycles.

In terms of the maximum flux density of a BHB, $F_{\rm x,max}$
(Table~\ref{tab:bhb1}), the 2--11~keV luminosity is: $L_{\rm x}/L_{\rm
Edd} \approx~2.6\times10^{35} \times F_{\rm x,max}(\mu$Jy) $\times$
($D$/10~kpc)$^2$ (Bradt \& McClintock 1983).  For a Crab--like
spectrum (\S4.1.2), the flux in the 0.5--10~keV (ULX) band is a factor
of 1.9 greater (although this is somewhat of an overestimate compared
to the use of a thermal spectrum).  Including this factor and using
the distances and peak fluxes in Table~\ref{tab:bhb1}, we find that
the three most luminous BHBs are V4641~Sgr
($6.2 \times 10^{39}$~\ergsec), 4U~1543--47
($4.2 \times 10^{39}$~\ergsec) and GRS~1915+105 ($2.4
\times 10^{39}$~\ergsec).  Thus, at peak luminosity these three BHBs
appear to be in the same league as many of the ULXs observed in
external galaxies.  Moreover, using the mass measurements from
Table~\ref{tab:bhb2}, it appears that all three BHBs were
super--Eddington at maximum (0.5--10~keV): $L_{\rm x}/L_{\rm
Edd}$~=~7.0, 3.5 and 1.4 for V4641~Sgr, 4U~1543--47 and GRS~1915+105,
respectively.

This comparison of BHBs and ULXs is somewhat problematic: First, the
distances of the BHBs are uncertain, and we have no direct
measurements of their fluxes in the 0.5--2~keV band.  Nevertheless,
the results quoted above suggest that the peak luminosities of a few
BHBs approach the peak luminosities observed for ULXs to within a
factor of $\approx5$ (e.g., Miller et al. 2003b).  Second, the
luminosity shortfall of BHBs may be ascribable to the small sample of
18 BHBs compared to the much larger sample of comparable systems that
have likely been detected in surveys of external galaxies.  Finally,
some ULXs exhibit spectral properties unlike those of BHBs. In
particular, the cool disk spectra and the longevity at high luminosity
of the most luminous ULXs may distinguish them from BHBs in the Milky
Way (Miller et al. 2003a; 2003b).  The differences among ULX spectra
also suggest that the sample may be heterogeneous.

We conclude by noting that super--Eddington luminosities have plainly
been observed for a few NS systems.  The most clear--cut case is
A0535--668, the ``LMC transient.''  This pulsating NS binary with a
firm distance of D = 50~kpc (Freedman et al. 2001) achieved a peak
luminosity of $L_{\rm x}~\approx 1.2 \times 10^{39}$~\ergsec, assuming
isotropic emission (Bradt \& McClintock 1983, and references therein).
This is 6.9 times the Eddington luminosity of a canonical 1.4~\msun~NS
or 3.8 times the Eddington luminosity of a hypothetical 2.5~\msun~NS.


\section{Emission states of black hole binaries}

As discussed in \S4.1 and \S4..2, BHBs exhibit thermal and nonthermal
components of X--ray emission, both of which can vary widely in
intensity.  It has long been recognized that BHBs undergo transitions
between quasi--stable states in which one or the other of these
components may dominate the X--ray luminosity.  In the past, the study
of BH emission states was based almost exclusively on X--ray spectral
and timing studies.  More recently, however, the results of X--ray
studies have been supplemented with critical contributions by radio,
optical and gamma--ray observers to give us a more physical and
fruitful framework for regarding the emission states.  In the
following sections, we review the characteristic behavior that defines
each of the principal X--ray states of BHBs.  We then describe the
current picture of each state in terms of physical structures and the
nature of the accretion flow.  Finally, we discuss the prospects of
using these states to deduce the properties of BHs.

\subsection{Historical notes on X--ray states}

In the spring of 1971, Tananbaum et al. (1972) observed a remarkable
X--ray state change in Cyg X--1 during which the average soft flux
(2--6~keV) decreased by a factor of 4 and the average hard flux
increased by a factor of 2.  Simultaneously, the radio counterpart of
Cyg~X--1 brightened.  It was later found that luminous X--ray novae
such as A0620--00 exhibited similar spectral transitions, suggesting
that common emission mechanisms were at work in both persistent and
transient BHCs (Coe et al. 1976).  These
early results suggested that such global spectral changes might
signify important changes in accretion physics.

As the many light curves in \S4.2 illustrate, the soft X--ray state is
generally seen at higher luminosity, motivating frequent references to
the {\it high/soft} (HS) state. In this state, the spectrum may also
display a ``hard tail'' that contributes a small percentage of the
total flux.  As shown by the synoptic studies discussed in \S4.2.2,
the soft state is best explained as $\sim$1 keV thermal emission from
a multi--temperature accretion disk (see \S4.1.5), as foreseen in the
standard theory for accretion in BHBs (Shakura \& Sunyaev 1973).
However, it has been found that the soft state of Cyg X--1 is not
consistent with a thermal interpretation (Zhang et al. 1997b), and this
has caused considerable confusion as to the proper way to understand
Cyg X--1 and/or describe the HS state.  In seeking a physical basis
for describing X--ray states, it turns out that Cyg~X--1 is not a good
choice as a prototype, and further remarks about the states in Cyg
X--1 are given in a separate section below (\S4.3.9).

In the hard state the 2--10~keV intensity is comparatively low,
prompting the name {\it low/hard} (LH) state.  The spectrum
is nonthermal and conforms to a power--law with a typical photon index
$\Gamma \sim$ 1.7 (2--20~keV).  In this state, the disk is either not
detected at 2--10 keV (e.g., Belloni et al. 1999), or it appears much
cooler and larger than it does in the soft state (Wilms et al. 1999;
McClintock et al. 2001b).

An additional X--ray state of BHBs was identified in the {\it Ginga}
era (Miyamoto et al. 1993).  It is characterized by the appearance of
QPOs in the presence of both disk and power--law components, each of
which contributes substantial luminosity (e.g., $> 0.1 L_{\rm Edd}$;
van der Klis 1995).  In this state, which is referred to as the {\it
very high} (VH) state, the power--law component is observed to be
steep ($\Gamma \sim$ 2.5).  Initially, there were only two BHBs
(GX339--4 and Nova Mus 1991) that displayed this behavior, but many
additional examples have been seen in the {\it RXTE} era.

It was first thought that the two nonthermal states (i.e., LH and VH
states) could be distinguished through differences
in their photon spectral indices, luminosities, and power density
spectra.  However, as shown below, the latter two differences have
become blurred; nevertheless, the spectral index continues to be a
valid discriminator.  The importance of distinguishing between the LH
and VH states was emphasized in a $\sim$40--500~keV study of seven
BHBs with the OSSE instrument aboard the {\it Compton Gamma Ray
Observatory} (Grove et al. 1998).  The gamma--ray spectra of these
sources separate naturally into two distinct groups which 
correspond to the LH~state and the VH~state, respectively: (1) For
the first group it was shown that the X--ray LH~state corresponds with
a ``breaking gamma--ray state'' in which the spectrum below
$\sim$100~keV is harder than that of the VH~state, but then suffers an
exponential cutoff near 100~keV. (2) The second group exhibits a
power--law gamma--ray spectrum with photon index $2.5 < \Gamma < 3.0$
over the entire range of statistically significant measurements. This
gamma--ray photon index is consistent with the X--ray photon index of
the VH~state.  Furthermore, contemporaneous X--ray observations (e.g.,
with {\it ASCA}) confirmed that a luminous thermal component coexists
with the power--law component, which is one characteristic of the VH
state noted above.

\subsection{X--ray states as different physical accretion systems} 

Several recent developments in the study of BHBs has taken us beyond a
largely phenomenological description of X--ray states to one based on
physical elements (e.g., accretion disk, ADAF, jet, and corona).
Although this work is still incomplete, the fundamental distinctions
between the states are becoming clearer.  For example, a key
development in this regard is the recognition of a persistent radio
jet associated with the LH state that switches off when the source
returns to the HS state (\S4.3.6; Ch. 9).  Another example, revealed
by gamma--ray observations, is the very different coronal structure
that is responsible for the clear--cut distinction between the
LH~state and the VH~state (\S4.3.1).  Arguably, each X--ray state can
be regarded as a different accretion system that can be used in unique
ways to study accretion physics and the properties of accreting BHs.

In the sections below, we review each of the four canonical X--ray
states of BHBs, including the long--lived {\it quiescent} state.  We
illustrate the uniform X--ray properties of each of the three active
states by showing X--ray spectra and power spectra for several BHBs
and BHCs observed by {\it RXTE}.  We also examine a possible fifth
X--ray state, the {\it intermediate} state, as part of our discussion
of the VH~state.  While presenting this overview, we suggest an
alternative set of state names that are motivated by the kinds of
emergent physical pictures mentioned above.  Although the new state
names depend critically on multiwavelength results (i.e., radio to
gamma--ray), we nevertheless attempt to define them on the basis of
X--ray data.  Furthermore, based on extensive observations of many
sources with {\it RXTE} (e.g., see \S4.2.2), {\it we abandon
luminosity as a criterion for defining the states of BHBs} (with the
exception of the {\it quiescent} state). We do not deny that there are
correlations between states and luminosity in many sources, in
particular the tendency of the HS to occur at higher luminosity
compared with the hard state. However, as shown below, there are clear
exceptions to these trends and each X-ray state has now been observed
to span a range of two or more decades in X-ray luminosity.


\subsection{Notes on X--ray spectral analyses}

Many spectral models have been developed to describe one or more of
the spectral states of BHBs.  Most models for the nonthermal continuum
components invoke inverse Compton or synchrotron emission, but these
mechanisms can be applied with many different assumptions and
geometric details. Some models closely constrain the relationship
between spectral components (e.g., thermal emission providing seed
photons for Comptonization), while others allow the spectral
components to vary independently.  In presenting this review, we have
adopted the following pragmatic and generic strategy.  As discussed in
\S4.2.2, the spectra of BHBs are well described by a model consisting
of a multi--temperature accretion disk component and a power--law
component (which may require an exponential cutoff at high energy).
This model provides a robust, first--order description of BHB spectra
that covers all of the emission states, and we adopt it to help define
the states and to compare the luminosities of the thermal and
nonthermal components.  We also include additional features in the
model, such as Fe line emission and a disk reflection component, when
the normalization parameter for such a feature is significant at the
level of $5~\sigma$.  In the following discussions of each X--ray
state, we comment on some physical interpretations and controversies
related to the problem of determining the origin of the power--law
component.


\subsection{Quiescent state} 

A BHB spends most of its life in a {\it quiescent} state that can be
summarized as {\it an extraordinarily faint state ($L_x =
10^{30.5}-10^{33.5}$ \ergsec), with a spectrum that is distinctly
nonthermal and hard ($\Gamma = 1.5-2.1$)}.  The first short--period
X--ray nova to be detected in quiescence was A0620-00 ($P_{\rm
orb}~=7.8$~hr); its X--ray luminosity was several times
$10^{30}$~\ergsec, which is only $\sim10^{-8}$ of its outburst
luminosity (McClintock et al. 1995; Narayan et al. 1996).  The
long--period systems, however, are significantly more luminous in
quiescence because their mass transfer rates are driven by the nuclear
evolution of their secondaries rather than by gravitational radiation
(Menou et al. 1999).  For example, the X--ray luminosity of V404~Cyg
($P_{\rm orb}~=155.3$~hr) is typically $L_{\rm
x}~\sim~10^{33}$~\ergsec~(Kong et al. 2002), but can vary by an order
of magnitude in one day (Wagner et al. 1994).

It is now possible to make sensitive measurements in the {\it
quiescent} state using {\it Chandra} and {\it XMM--Newton}.  The
minimum quiescent--state luminosities (0.5--10~keV) of five BHs and
stringent upper limits on two others have been reported by Garcia et
al. (2001) and Narayan et al. (2002). Three additional BHBs were
observed in quiescence more recently (Sutaria et al. 2002; Hameury et
al. 2003; McClintock et al. 2003a).  Considering only the five
short--period systems ($P_{\rm orb}~\lesssim~1$~day; see Narayan et
al. 2002) that have been detected, one finds that four of them
(XTE~J1118+480, GRO~J0422+32, GS 2000+25, and A0620--00) have
Eddington--scaled luminosities that are 
$\approx~10^{-8.5}$. GS~1124--68 is more luminous by about an
order--of--magnitude.  As Garcia et al. (2001) and Narayan et
al. (2002) show, the Eddington--scaled luminosities of several
ostensibly similar X--ray novae that contain NS primaries are about
100 times higher, a conclusion that is quite robust.  In the context
of the advection--dominated accretion flow model, these authors argue
that the relative faintness of the BH X--ray novae provides strong
evidence that they possess event horizons.  For a thorough discussion
of their model and several alternative models, see Narayan et
al. (2002).  Apart from any specific model, it appears quite
reasonable to suppose that the established faintness of quiescent BHs
is somehow connected with the existence of the event horizon.

The quiescent spectra of BHBs are well fitted by a single power--law
plus interstellar absorption.  The best--determined photon spectral
indices are consistent with the value
$\Gamma~\approx~2$. Specifically, for A0620--00 and XTE~J1118+480,
respectively, one has $\Gamma~=~2.07(+0.28,-0.19)$ (Kong et al. 2002)
and $\Gamma~=~2.02\pm 0.16$ (McClintock et al. 2003a), where the
column density is determined from the optical extinction.  Only V404
Cyg allows a useful determination of both the column density and the
photon index: $N_{\rm H} = (6.98 \pm 0.76) \times 10^{21}$ cm$^{-2}$
and $\Gamma = 1.81 \pm 0.14$ (Kong et al. 2002).
    
\begin{figure}
\epsfxsize20pc 
\epsfysize20pc 
\hspace*{-0.13in}
\epsfbox{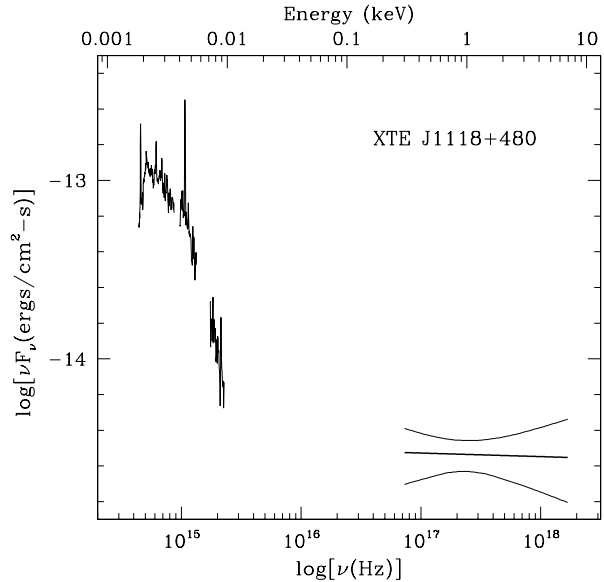}
\vspace*{-0.20in}
\caption{Spectrum of XTE~J1118+480 in the {\it quiescent} state based on
simultaneous, multiwavelength observations.  The optical spectrum of
the mid--K dwarf secondary has been subtracted.  Note the Planckian
shape of the optical/UV continuum, which is punctuated by a dominant
Mg~II 2800\AA\ line and a strong H$\alpha$ line on the far left.  The
best--fit X--ray model is indicated by the heavy, horizontal line; the
90\% error box is defined by the flanking curved lines.}
\label{mcclin}
\vspace*{-0.07in}
\end{figure}

A multiwavelength spectrum of XTE~J1118+480 in quiescence is shown in
Figure~\ref{mcclin} (McClintock et al. 2003a).  This shortest--period
BHB ($P_{\rm orb}~=~4.1$~hr) is located at b~=~62$^{\rm o}$, where the
transmission of the ISM is very high (e.g., 70\% at 0.3 keV).  A very
similar multiwavelength spectrum was observed earlier for A0620--00
(McClintock \& Remillard 2000), which implies that the spectrum shown
in Figure~\ref{mcclin} represents the canonical spectrum of a BHB
radiating at $10^{-8.5}L_{\rm Edd}$.  The spectrum is comprised of two
apparently disjoint components: a hard X--ray spectrum with a photon
index $\Gamma~=~2.02~\pm~0.16$, and an optical/UV continuum that
resembles a 13,000~K disk blackbody spectrum punctuated by several
strong emission lines.

The ADAF/disk model (see \S4.1.5) accounts well for the following
properties of BHBs in the {\it quiescent} state: (1) The hard
power--law spectra (Narayan et al. 1996; Narayan et al. 1997; Hameury
et al. 1997; Quataert \& Narayan 1999; McClintock et al. 2003a); (2)
the faintness of BHs relative to NSs (Narayan et al. 1997; Garcia et
al. 2001; Narayan et al. 2002; (3) the several--day delay in the
optical/UV light curve when X--ray novae go into outburst (Hameury et
al. 1997); and (4) the broadband spectrum shown in Figure~\ref{mcclin}
(McClintock et al. 2003a).  Especially significant is the prediction,
confirmed by observations, that the accretion disk is truncated at a
large inner radius in both the {\it quiescent} and LH states (Narayan
1996; Esin et al. 1997; McClintock et al. 2001b; McClintock et
al. 2003a).
 

\subsection{Thermal--dominant (TD) state or high/soft (HS) state}

\begin{figure}
\epsfxsize33.14pc 
\epsfysize33.14pc 
\hspace*{0.03in}
\hspace*{-0.17in}
\epsfbox{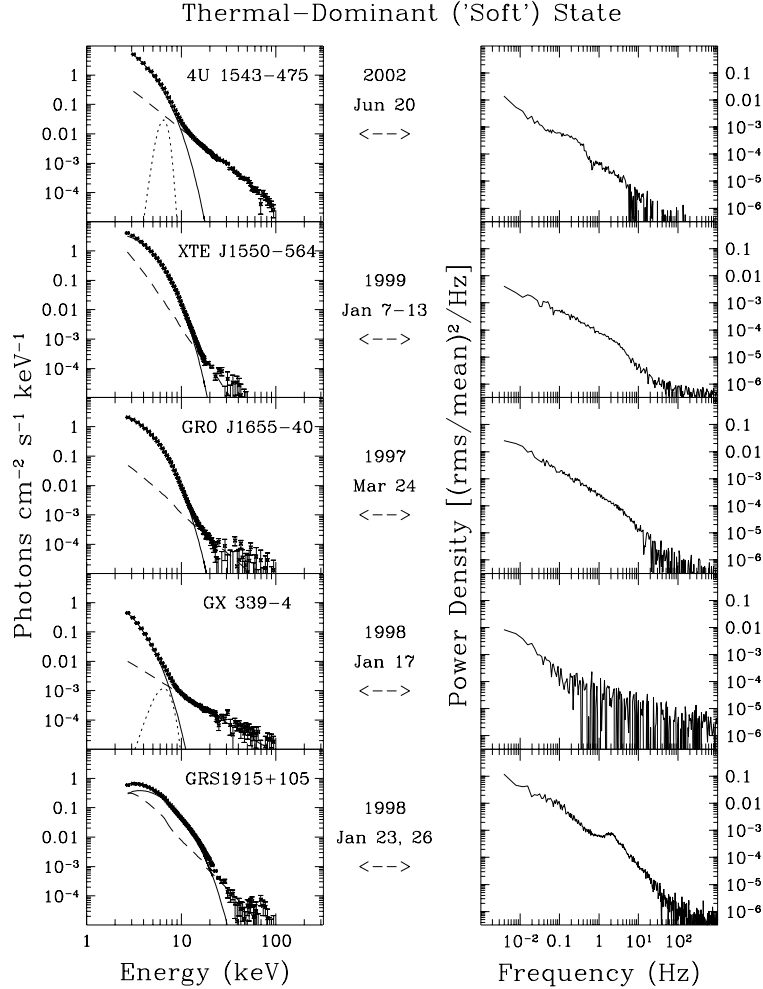}
\vspace*{-0.28in}
\caption{Sample X--ray spectra of BHBs in the X--ray state for which the
dominant component is thermal emission from the accretion disk.  The
energy spectra (left) are decomposed into a thermal component, which
dominates below $\sim$10~keV (solid line), and a faint power--law
component (dashed line); GRS~1915+105 is modeled with a cutoff
power--law (see text).  For two of the BHBs, an Fe line component is
included in the model (dotted line). The corresponding power spectra
are shown in the panels on the right.}
\label{spec-td1}
\vspace*{-0.07in}
\end{figure}

\begin{figure}
\epsfxsize33.14pc 
\epsfysize33.14pc 
\hspace*{0.03in}
\hspace*{-0.17in}
\epsfbox{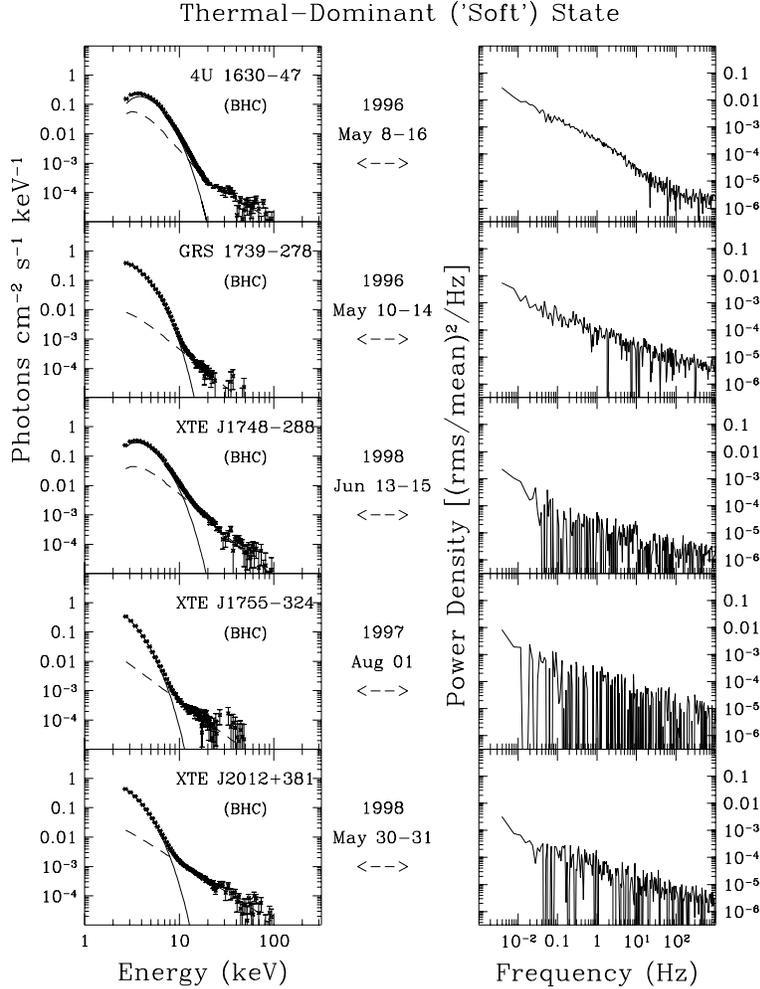}
\vspace*{-0.28in}
\caption{{\it Left panels:} Sample X--ray spectra of BHCs in the
TD X--ray state. In the panels on the left, the energy
spectra are shown deconvolved into a thermal component due to the
accretion disk (solid line) and a power--law component (dashed line).
The corresponding power spectra are shown in the right half of the
figure.}
\label{spec-td2}
\vspace*{-0.07in}
\end{figure}

As discussed in \S4.1.5, considerations of basic principles of physics
predict that accreting BHs should radiate thermal emission from the
inner accretion disk (Shakura \& Sunyaev 1973).  It was therefore
readily accepted that the soft X--ray state of BHBs represents thermal
emission. Confirmations of this picture are largely based on the
successful ability to describe the soft X--ray component using the
simple multi--temperature accretion disk model (MCD model; \S4.1.5).
In Figures \ref{spec-td1} \& \ref{spec-td2} we show the energy spectra
and power spectra, respectively, of 10 BHBs in the
``thermal--dominant'' (TD) or {\it high/soft} (HS) X--ray state as
observed by {\it RXTE}.  The thermal component of the model, where it
can be distinguished from the data, is shown as a solid line, and the
power--law component is shown as a dashed line.  Typically, below
about 10~keV the thermal component is dominant.  With a few
exceptions, the temperature of this component is in the range
0.7--1.5 keV (Table~\ref{tab:spfit}). The power--law component is
steep ($\Gamma = 2.1-4.8$) and faint.  In GRS~1915+105 the power--law
falls off even more steeply with an e--folding cutoff energy of
3.5~keV; similar behavior has been reported for GRO~J1655--40 (Sobczak
et al. 1999).

\begin{figure}
\epsfxsize58.0pc
\epsfysize58.0pc
\hspace*{-0.31in}
\hspace*{-0.18in}
\hspace*{0.31in}
\epsfig{file=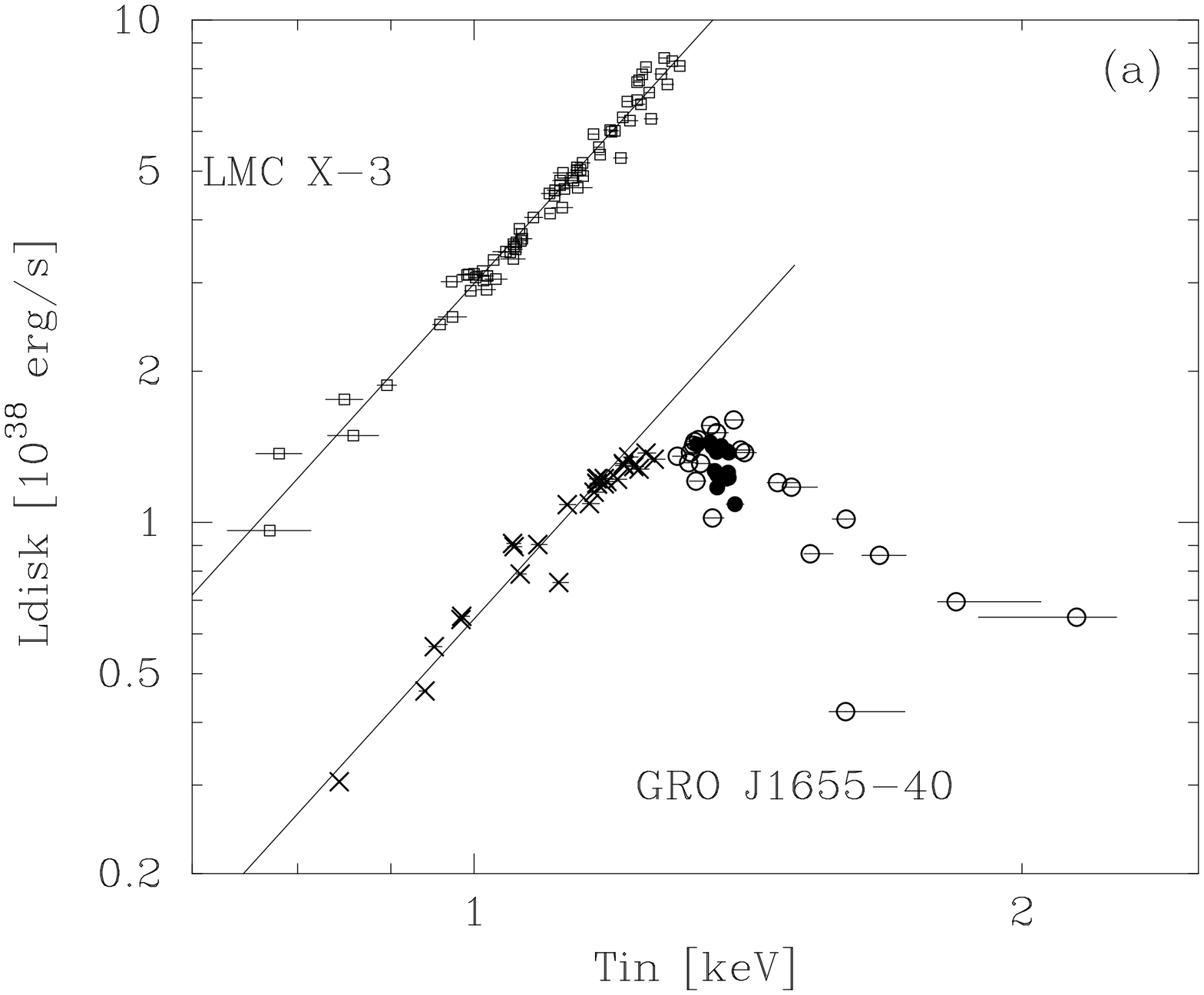,height=22pc,angle=-90}
\vspace*{-0.08in}
\caption{The accretion disk luminosity {\it vs} temperature at
the inner accretion disk (Kubota et al. 2001).  For GRO J1655--40,
the symbol type denotes the time periods: early outburst (filled
circles), first phase (open circles), and second phase (crosses). The
solid lines represent the $L_{\rm disk}\propto T_{\rm in}^4$ relation.}
\label{kubo}
\vspace*{-0.07in}
\end{figure}

In Figure~\ref{spec-td1} we feature data for BHBs obtained during
pointed observations with {\it RXTE}, while data for BHCs are shown in
Figure~\ref{spec-td2}.  The spectra of four of the sources in the
figures (4U~1543--47, GX~339--4, XTE~J1755--324, and XTE~J2012+381)
correspond to the maximum 2--20 keV luminosities observed during {\it RXTE}
pointings (i.e., considering all possible states).  The power density
spectra (PDS) in Figures \ref{spec-td1} \& \ref{spec-td2} show that
the variability in the TD state is either weak or the power scales
roughly as $\nu^{-1}$, which is a characteristic of many physical
processes including turbulence (Mandelbrot 1999).  The total rms power
($r$) integrated over 0.1--10 Hz in these PDS is in the range
$0.01~\lesssim r~\lesssim 0.06$, which is significantly below that of
the LH state. QPOs in the range 0.1--30 Hz are generally not seen in
individual TD observations, but large groups of PDS in the TD state
have yielded weak QPOs in two cases.  A 0.3\% (rms) QPO at 27 Hz was
seen in a sum of 27 observations (``1997 soft state'') of
GRO~J1655-40 (Remillard et al. 1999), and a similar (0.3\%) QPO at 17
Hz was seen in 69 observations of XTE~J1550-564 in the soft state
during 1998--1999 (MJD 51160--51237; Homan et al. 2001).

In the following section, we show that a transition to the LH~state is
followed by the appearance of a hard and dominant power--law spectrum,
while the accretion disk, if visible, shows a substantial decrease in
temperature.  On the other hand, a transition to the VH state
(\S4.3.7) is marked by a steeper power--law spectrum accompanied by
either a normal ($\sim 1$ keV) disk or one that appears hot with a
small inner radius.  This latter transition is also accompanied by the
presence of QPOs that appear when the disk contribution to the total,
unabsorbed flux at 2--20 keV falls below the level of 0.75 (Sobczak et
al. 2000a).  {\it We thus define the TD state as the set of conditions
for which the disk--flux fraction is above 75\% (2--20 keV), the PDS
shows no QPOs or very weak features ($rms << 1$\%), and the power
continuum is also weak : $r~\lesssim~0.06$ integrated over 0.1--10
Hz.}

As further support for a thermal interpretation of the soft X--ray
component, many studies have found evidence for disk luminosity
variations in which the inner disk radius (which scales as the square
root of the MCD normalization parameter) appears constant while the
luminosity variations depend only on changes in temperature (\S4.2.2).
This effect is reported in a study of LMC X--3 (Kubota et al. 2001)
and is illustrated in Figure~\ref{kubo}. The measured disk flux and
apparent temperature successfully track the relation $L ~ \propto ~
T^4$ (solid line) expected for a constant inner disk radius.  The
figure also shows gross deviations from this relation associated with
the VH~state of GRO J1655--40, a topic that is addressed below in
\S4.3.7.

Because of the successes of the simple MCD model, the inner disk
radius has been used to provide a type of spectroscopic parallax.  In
principle, the inner disk radius can be deduced for those sources for
which the distance and disk inclination are well constrained from the
disk normalization parameter, ($R_{\rm in}$/D)$^2$~cos$\theta$,
where $R_{\rm in}$ is the inner disk radius in kilometers, $D$ is the
distance to the source in units of 10~kpc, and $\theta$ is the
inclination angle of the system (e.g., Arnaud \& Dorman 2002).
However, the MCD model is Newtonian, and the effects of GR and
radiative transfer need to be considered.  GR predicts a transition
from azimuthal to radial accretion flow near $R_{\rm ISCO}$, which
depends only on mass and spin and ranges from $1-6$\Rg~for a prograde
disk (\S4.1.5).  Therefore, for a system with a known distance and
inclination, in principle it may be possible to estimate the spin
parameter via an X--ray measurement of the inner radius and an optical
determination of the mass.  In lieu of a fully relativistic MHD model
for the accretion disk, one could attempt to correct the MCD model
parameters to account for the effects of radiative transfer through
the disk atmosphere (Shimura \& Takahara 1995) and for modifications
on the structure and emissivity of the inner disk due to GR (Zhang et
al. 1997a).  However, the accuracy of these corrections has been
challenged (Merloni et al.  2000; Gierlinski et al. 1999), and at
present the chief value of the MCD model is in monitoring the
temperature and the fractional flux contribution from the accretion disk.


\subsection{Hard X--ray state associated with a 
steady radio jet}

The conventional name for this state is the {\it low/hard} (LH) state;
however, henceforth we refer to it simply as the {\it hard} state for
the reasons given in \S4.3.2.  As noted earlier (\S4.3.1), Cyg~X--1 and
many transient sources have been observed to undergo transitions to a
hard, nonthermal X--ray spectral state. This usually occurs at
luminosities below that of the TD state, and the spectrum can be
modeled (e.g., 1--20 keV) as a power--law function with a photon index
$\sim$1.7.  In some sources, such as Cyg~X--1 and GS~1354--64, this
{\it hard} state is accompanied by a broad enhancement at 20--100 keV
which is interpreted as reflection of the power--law component from
the surface of the inner accretion disk (Di Salvo et al. 2001). This
component is discussed further at the end of this section.

In recent years there have been rapid advances in associating the
X--ray {\it hard} state with the presence of a compact and
quasi--steady radio jet (for a thorough review, see Ch. 9).  This
relationship, which constitutes one of the foundations of the ``disk :
jet'' connection, is based on at least three arguments.  First, VLBI
radio images have shown a spatially--resolved radio jet during
episodes of quasi--steady radio and hard X--ray emission from
GRS~1915+105 (Dhawan et al. 2000) and Cyg X--1 (Stirling et
al. 2001). In both instances the radio spectrum was flat or inverted.
Second, more generally (i.e., when VLBI images are not available),
X--ray sources that remain in the {\it hard} state for prolonged
periods (weeks to years) are highly likely to show correlated X--ray
and radio intensities and a flat radio spectrum (Fender et al. 1999a;
Fender 2001; Corbel et al. 2000; Klein--Wolt et al. 2002; Marti et
al. 2002; Corbel et al. 2003).  In one of these examples, GX~339--4,
the jet interpretation is further supported by the detection of 2\%
linear radio polarization with a nearly constant position angle
(Corbel et al. 2000).  Finally, it is now routine to witness the
quenching of the persistent radio emission whenever an X--ray source
exits the {\it hard} state and returns to the TD state (Fender et
al. 1999b; Brocksopp et al. 1999; Corbel et al. 2000).

\begin{figure}
\epsfxsize33.14pc 
\epsfysize33.14pc 
\hspace*{0.03in}
\hspace*{-0.17in}
\epsfbox{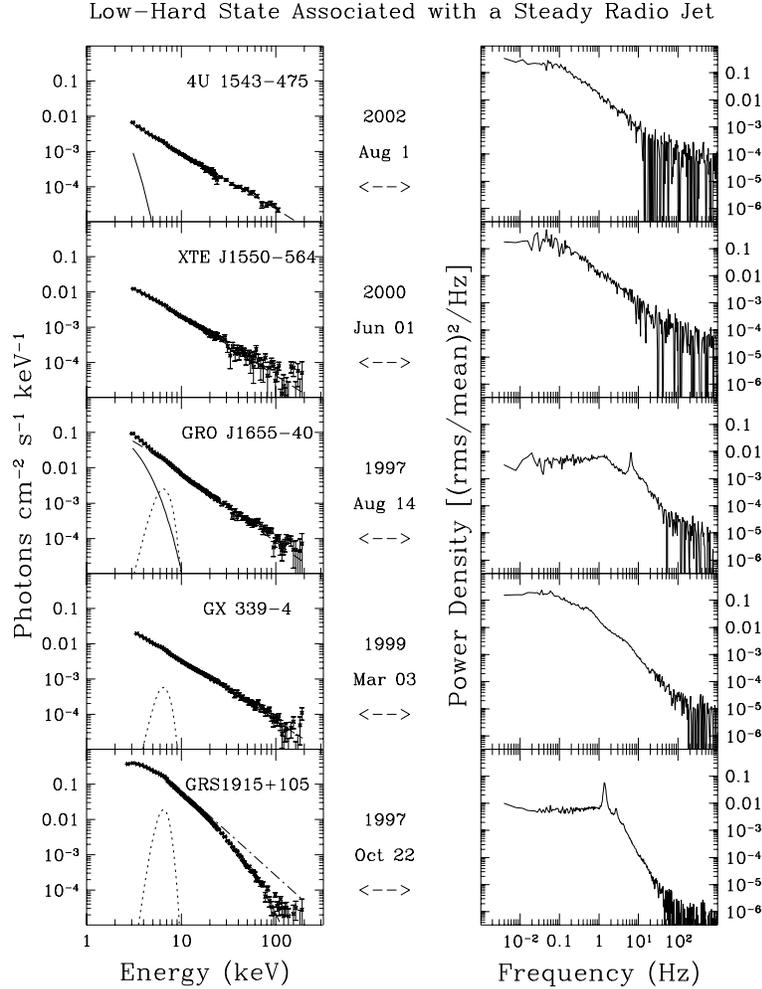}
\vspace*{-0.24in}
\caption{Sample spectra of BHBs in the {\it hard} state.  The
energy spectra are characterized by a relatively flat power--law
component that dominates the spectrum above 1 keV. A second
characteristic of the {\it hard} state is the elevated continuum
power in the PDS. This state is associated with the presence of a
steady type of radio jet (see text). The selected X--ray sources are
the same BHBs shown in Figure~\ref{spec-td1}.  The individual
spectral components include the power--law (dashed line) and, if
detected, the accretion disk (dotted line) and a reflection component
(long dashes).}
\label{spec-lh1}
\vspace*{-0.07in}
\end{figure}

In Figure~\ref{spec-lh1}, the five BHBs shown in the {\it hard} state
are the same sources shown earlier in the TD state
(Fig.~\ref{spec-td1}).  In the following cases, the X--ray data have
been selected to coincide with specific radio observations:
Flat--spectrum radio emission was reported for both XTE J1550--564 on
2000 June 1 during the decay of its second outburst (Corbel et
al. 2001) and for GX 339--4 on 1999 March 3 (Corbel et al. 2000).  In
addition, both the radio and X--ray emission of GRS~1915+105 are
fairly steady on 1997 October 22, which coincides with one of the days
in which the core radio image shows the extended structure of a
nuclear jet (Dhawan et al. 2000).
  
\begin{figure}
\epsfxsize33.14pc 
\epsfysize33.14pc 
\hspace*{0.03in}
\hspace*{-0.17in}
\epsfbox{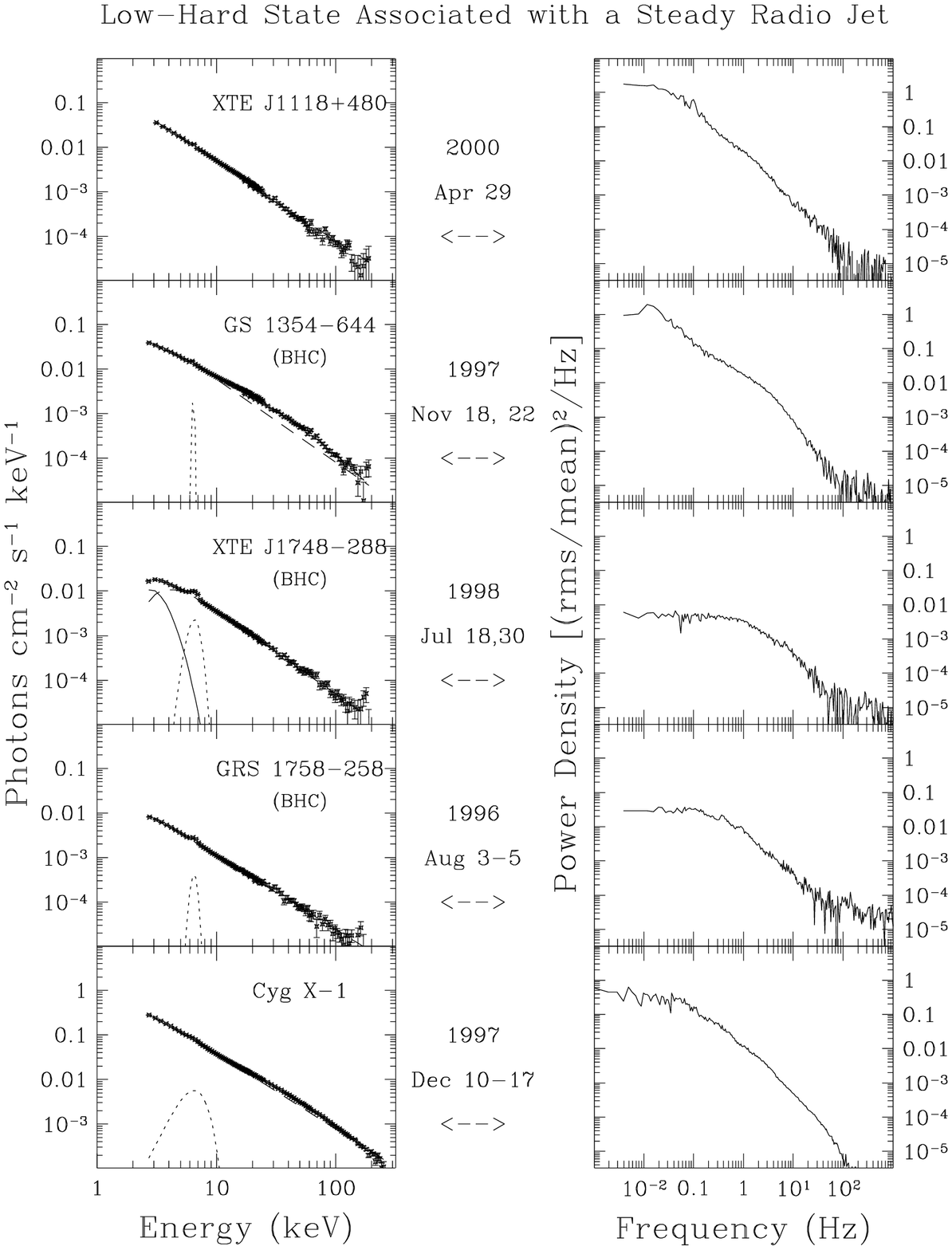}
\vspace*{-0.24in}
\caption{A second sample of BHBs or BHCs seen in the {\it hard} state.
The observations of XTE~J1748--288 occurred during outburst decay,
while XTE~J1118+480 and GS~1354--64 are seen in the {\it hard} state
at the peaks of their respective outbursts.  GRS~1758--258 and Cyg
X--1 spend most of their time in the {\it hard} state. There is radio
coverage that confirms the presence of a flat radio spectrum for all
the sources except XTE~J1118+480.  The line types denoting the
spectral components follow the convention of Figure~\ref{spec-lh1}}.
\label{spec-lh2}
\vspace*{-0.07in}
\end{figure}

A second sample of BHBs and BHCs in the {\it hard} state is shown in
Figure~\ref{spec-lh2}. The {\it hard} state for XTE~J1748--288
occurred during decay from the HS state (Revnivtsev et al. 2000c). On
the other hand, the data shown for XTE~J1118+480 and GS~1354--64
correspond with outburst maxima; these outbursts never reached the HS
state (Revnivtsev et al. 2000a; Frontera et al. 2001b). Associated
radio emission for these two sources was reported with jet
interpretations by Fender et al. (2001) and Brocksopp et al. (2001),
respectively. Finally, GRS~1758--258 and Cyg~X--1 spend most of their
time in the {\it hard} state, and their energy spectra are shown in
the bottom two panels of Figure~\ref{spec-lh2}.  These {\it RXTE}
observations happened to coincide with radio observations that confirm
a flat radio spectrum. In the case of GRS~1758--258, core radio
emission (as distinct from the extended radio lobes observed for this
source) was reported on 1998 August 3 and 5 by Marti et al. (2002),
and a clear detection of Cyg X--1 during 1997 December 12--17 is
evident in the public archive of the Greenbank Interferometer
available on the NRAO web site.

The physical condition of the accretion disk in the {\it hard} state
is a subject of great significance in the effort to build a detailed
physical model for both the jet and the X--ray source. Observations
with {\it ASCA}, which provided sensitivity in the range 0.5 to 9 keV,
showed that the {\it hard} states of both GX~339--4 (Wilms et
al. 1999) and Cyg~X--1 (Takahashi et al. 2001) exhibit power--law
spectra with an additional soft X--ray excess that can be modeled as a
large and cool ($\sim$0.1--0.2 keV) accretion disk.  The spectral
decompositions illustrated in Figure~\ref{spec-lh1} provide some
evidence for a soft disk component in both GRO~J1655--40 and
GRS~1915+105, although {\it RXTE} is much less sensitive than {\it
ASCA} to thermal spectra with temperatures well below 1~keV.  By far,
the best direct measurements of the temperature and inner radius of an
accretion disk in the {\it hard} state have been made for
XTE~J1118+480, which has an extraordinarily small interstellar
attenuation (e.g., only 30\% at 0.3 keV).  Based on simultaneous {\it
HST}, {\it EUVE} and {\it Chandra} observations made in outburst, it
was determined that the inner disk radius and temperature for the MCD
model were $\gtrsim$100~\Rg~and $\approx~0.024$~keV, respectively
(McClintock et al. 2001b).  Somewhat higher temperatures
($\approx$~0.035--0.052~keV) have been inferred from observations
using BeppoSAX (Frontera et al. 2003).

While it seems clear that the blackbody radiation appears truncated at
a large radius ($\sim$100~\Rg) in the {\it hard} state, the physical
state of the hot material within this large radius is still a matter
of debate.  Is the disk density truncated at this radius, as
envisioned in the ADAF model, or is a substantial amount of matter
present in a relativistic flow that is entrained in a jet (e.g.,
Markoff et al. 2001)?  Or is the inner disk basically intact and
either depleted of energy or veiled in some type of Compton corona?
For the latter possibility, the properties of the corona would be
strongly constrained by the absence of any normal ($\sim$1 keV)
thermal component in the {\it hard} state of XTE~J1118+480 (Esin et
al. 2001; Frontera et al. 2003).

Guidance in sorting out these options may eventually come from other
types of investigations, such as the study of correlated
optical/X--ray variability (Malzac et al. 2003).  Also promising are
spectral analyses that focus on broad Fe emission features (\S4.2.3)
or the X--ray reflection component (Done \& Nayakshin 2001).  These
spectral features depend on substantial density in the inner disk,
while the Fe line additionally reveals the pattern of Keplerian flow
modified by effects due to GR.  Systematic studies of these features
during different BHB states and transitions could help to determine
the physical changes in the inner disk associated with the {\it hard}
state. The reflection component is most apparent when the disk is
observed nearly face-on.  One such system is Cyg X--1, and a
reflection analysis has been reported by Done \& Zycki (1999).  They
find that the disk is physically truncated, but the transition radius
(tens of \Rg) is not as far from the event horizon as suggested by the
value of $R_{in}$ inferred from the disk spectrum in the {\it hard}
state (Takahashi et al. 2001).

The origin of the X--ray power--law is another aspect of the
controversy concerning the appropriate physical model for the {\it
hard} state.  As noted above, we regard the power--law fit as a
general signature of nonthermal radiation; however, the observed
spectrum can be produced by several different radiation
mechanisms. This is well illustrated in the case of XTE~J1118+480,
where the X--ray spectrum has been fitted by an ADAF model (Esin et
al. 2001), a synchrotron model (Markoff et al. 2001) and a
thermal--Comptonization model (Frontera et al. 2001b).

Despite the large uncertainties that remain for physical models, it
would appear that the association of the {\it hard} state with a
steady radio jet is an important step forward. And it does remain
possible to identify the {\it hard} state solely from X--ray spectral
and temporal properties, as had been done in the past (TL95). Using
Figures~\ref{spec-lh1} \& \ref{spec-lh2} and Table~\ref{tab:spfit},
{\it we conclude that the {\it hard} state is well characterized by
three conditions: the spectrum is dominated ($>$ 80\% at 2--20 keV) by
a power--law spectrum, the spectral index is in the range $1.5 <
\Gamma < 2.1$, and the integrated power continuum (0.1--10 Hz) is
strong and typically in the range $ 0.1 < r < 0.3$.}


\subsection{Steep power--law (SPL) state or very high (VH) state}

There are times when BHBs become exceedingly bright ($L_x > 0.2 L_{\rm
Edd}$), and the X--ray spectrum again displays substantial nonthermal
radiation, which may constitute 40--90\% of the total flux.  In such
cases the photon index is typically $\Gamma \ge 2.4$, which is steeper
than the index ($\Gamma \sim 1.7$) seen in the {\it hard} state. The
strength of this steep power--law component also coincides generally
with the onset of X--ray quasi--periodic oscillations (QPOs) in the
range 0.1--30 Hz.  This suite of characteristics was initially seen in
only two instances: during a bright outburst of GX 339--4 (Miyamoto \&
Kitamoto 1991) and near the time of maximum flux in X--ray Nova Muscae
1991 (= GS/GRS~1124--68; Miyamoto et al. 1993).  At the time, this
{\it very high} (VH) state was interpreted as a signature of the
highest rate of mass accretion in a BHB system (van der Klis 1995).

As mentioned previously (\S4.3.1), the high--energy spectra of several
BHCs observed with OSSE on {\it CGRO} (40--500 keV) reinforced the
distinction between the X--ray {\it hard} and VH states, showing that
the {\it hard}--state spectra exhibit a steep cutoff near 100 keV
(Grove et al. 1998).  On the other hand, the VH--state spectra showed
no evidence for a high--energy cutoff, while the photon index in the
X--ray and gamma--ray bands is the same ($\Gamma \sim$ 2.5--3.0).  The
unbroken power--law spectra observed by OSSE for five sources
suggested that these BHBs had been observed in the VH state, although
most of the observations were not accompanied by X--ray observations
that could assess the presence of QPOs.

The monitoring programs of {\it RXTE} have shown that the VH state is
both more common and more complicated than originally envisioned.
Some sources display both X--ray QPOs and a steep power--law component
at luminosity levels that are well {\it below} the maxima seen even in
their TD (HS) state (Remillard et al. 2002b; \S4.5). This topic is
considered further in \S4.3.8 below. In response to these
developments, we hereafter refer to this state as the ``steep
power--law'' state, or the SPL state, rather than the VH state.  We
adopt this new name because the steep power--law is a fundamental
property of this state, whereas a very high luminosity is not.  We
view the presence of QPOs as a confirming property of the SPL state.

\begin{figure}
\epsfxsize33.14pc 
\epsfysize33.14pc 
\hspace*{0.03in}
\hspace*{-0.17in}
\epsfbox{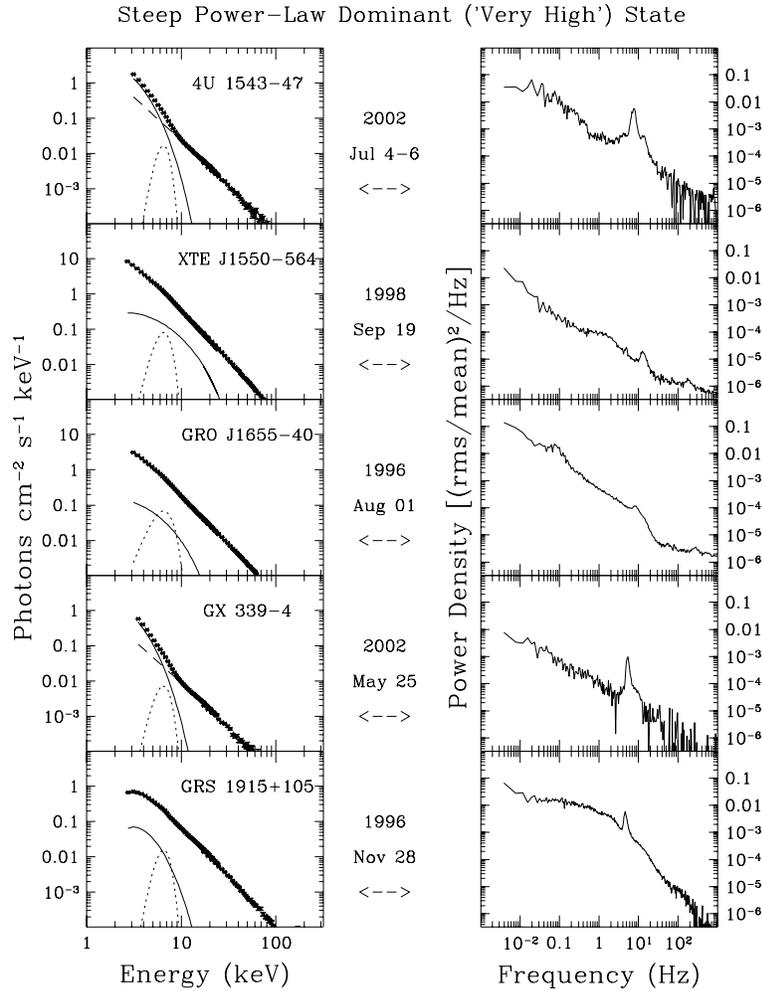}
 \vspace*{-0.24in}
\caption{X--ray spectra of BHBs in the SPL state, which is
characterized by a strong and steep power--law component in the energy
spectrum, along with the presence of X--ray QPOs. The dashed and dotted
lines follow the convention of earlier figures.}
\label{spec-spl1}
\vspace*{-0.07in}
\end{figure}

In Figure~\ref{spec-spl1} we show examples of the SPL state for the
same five BHBs considered in Figures~\ref{spec-td1} \& \ref{spec-lh1}.
The photon index of the steep power--law component
covers the range $2.4 < \Gamma < 3.0$ as shown in
Table~\ref{tab:spfit}, and QPOs are present with central frequencies
over the range 5--13 Hz .  The spectra of XTE~J1550--564 and
GRO~J1655--40 (Fig.~\ref{spec-spl1}) correspond to the highest
luminosities observed for these sources during pointed
observations with {\it RXTE} (Sobczak et al. 2000b; Sobczak et
al. 1999). Moreover, both observations revealed the presence of
high--frequency QPOs, 186~Hz and 300~Hz, respectively (Remillard et
al. 2002b).  The relationship between the SPL state and
HFQPOs will be discussed further in \S4.4.3.

\begin{figure}
\epsfxsize33.14pc 
\epsfysize33.14pc 
\hspace*{0.03in}
\hspace*{-0.17in}
\epsfbox{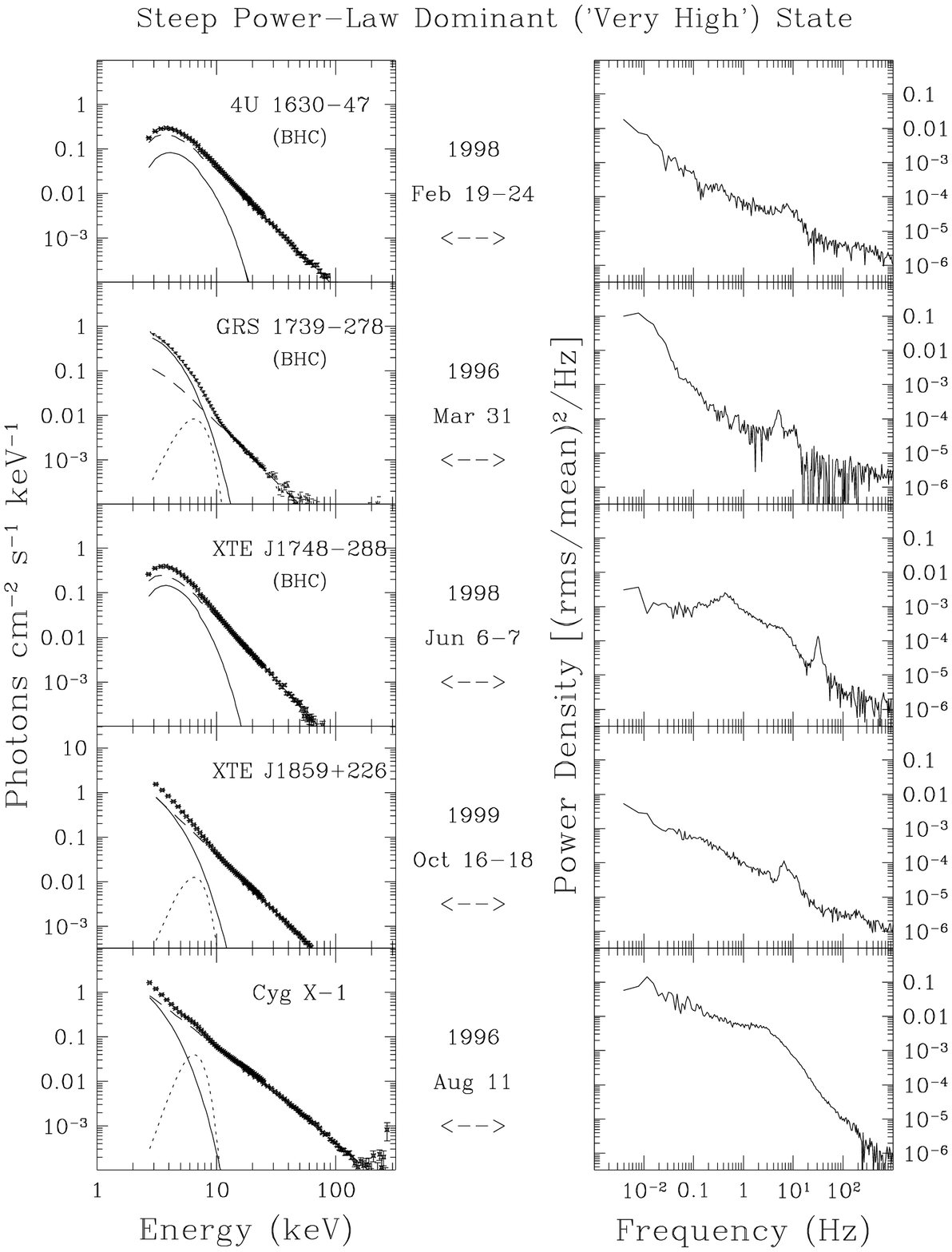}
\vspace*{-0.24in}
\caption{Additional examples of sources in the SPL state.  The
observations were selected near the times of global or local maxima in
the X--ray flux. Cyg X--1 is exceptional for its relatively low
luminosity and for the absence of QPOs (see \S4.3.9). The dashed and
dotted lines follow the convention of earlier figures.}
\label{spec-spl2}
\vspace*{-0.07in}
\end{figure}

In Figure~\ref{spec-spl2} we show spectra of the SPL state for three
BHCs and two additional BHBs.  For three of the sources we have
selected observations near the time of maximum flux, as seen in {\it
RXTE} pointed observations. For the two remaining sources we consider
local maxima: the 1996 soft--state episode of Cyg~X--1 and the second
outburst of 4U~1630--47. X--ray QPOs are seen in the top four panels
(5--8 Hz in three cases and 31~Hz in the case of XTE~J1748--288). The
power spectrum of Cyg~X--1 contains broad continuum features but no
X--ray QPOs, although the energy spectrum indicates that the source is
in an SPL--like state.

We can encompass all of these SPL examples with the following criteria
(see also Sobczak et al. 2000a).  {\it The SPL state is defined first
by the presence of a power--law component in the X--ray spectrum with
photon index $\Gamma > 2.4$. Secondly, either there are X--ray QPOs
present (0.1--30 Hz) while the power--law contributes more than 20\%
of the total (unabsorbed) flux at 2--20 keV, or the power--law
contributes more than 50\% of the total flux without detections of
QPOs.}

Transitions between the TD and {\it hard} states frequently pass
through intervals of the SPL state, and this has led some authors to
suggest that the SPL (or VH) state is itself an ``intermediate
state'' that lies between the TD and {\it hard} states (Rutledge et
al. 1999; Homan et al. 2001). However, this appears to be a radical
suggestion that gives inadequate weight to other SPL observations
associated with (1) the episodes of highest absolute luminosity in
many BHBs, (2) a distinct gamma--ray spectrum, and (3) occurrences of
QPOs at both high and low frequency. We therefore conclude that the
SPL state is a bona fide state of BHBs, and we note that nothing in
our state definitions constrains the order in which state transitions
should occur.

The radio properties of the SPL state are an important and complicated
topic that calls for the heightened attention of observers.  Here we
highlight several salient results.  The brightest days of
GRO~J1655--40 during its 1996--1997 outburst occurred in the SPL state
(1996 August) when the source appeared radio--quiet (e.g. Tomsick et
al. 1999). Similarly, Cyg X--1 becomes radio quiet whenever the
spectrum switches from the {\it hard} state to the SPL state. On the
other hand, the SPL state is also associated with the explosive
formation of radio jets.  For example, the giant X--ray flare in
XTE~J1550--564 (shown in Fig.~\ref{asm_bh} \& \ref{spec-spl1}) has
been linked to a relativistic mass ejection seen as a superluminal
separation of bipolar radio jets (Hannikainen et al. 2001). However,
there is a distinct possibility that the X--ray flare in the SPL state
may have occurred after the moment of ejection, at a time when the
radio properties of the core (i.e., the inner disk) are unknown.  This
conjecture is supported by radio observations of the same source
during the 2000 outburst when a radio flare was observed to decay
below detectable levels while the source remained in the SPL (VH)
state (Corbel et al. 2001). We conclude that the best available
evidence suggests that the SPL state is essentially radio quiet, while
the instability that causes impulsive jets is somehow associated with
the SPL state.

The physical origin of the SPL spectrum remains one of the outstanding
problems in high--energy astrophysics. The SPL spectrum extends to
$\sim$1~MeV in several sources: e.g., Cyg~X--1, GRO~J1655--40 and
GRO~J0422+32 (Table~\ref{tab:bhb2}). The spectrum may extend to even
higher energies, but present investigations are limited by photon
statistics. At stake is our understanding of accretion physics at the
extraordinary times of peak BHB luminosity.  Equally important is our
need to interpret the high--frequency QPOs associated with the
SPL state.  

Most models for the SPL state invoke inverse Compton scattering as the
operant radiation mechanism (e.g., Zdziarski 2000).  
The MeV photons suggest that the scattering
occurs in a nonthermal corona, which may be a simple slab operating on
seed photons from the underlying disk (Gierlinski et al. 1999;
Zdziarski et al. 2001).  Efforts to define the origin of the
Comptonizing electrons has led to more complicated geometric models
with feedback mechanisms, such as flare regions that erupt from
magnetic instabilities in the accretion disk (Poutanen \& Fabian
1999).  One early model, which was applied to AGN, invokes a strongly
magnetized disk and predicts power--law spectra extending to tens of
MeV (Field \& Rogers 1993).  An analysis of extensive {\it RXTE}
observations of GRO~J1655--40 and XTE~J1550--564 has shown that as
the power--law component becomes stronger and steeper, the disk
luminosity and radius appear to decrease while maintaining a high
temperature (see Fig.~\ref{kubo}). These results can be interpreted
as an observational confirmation of strong Comptonization of disk
photons in the SPL state (Kubota et al. 2001; Kubota \& Makishima
2003).

There are a number of alternative models for the SPL state. For
example, bulk motion Comptonization has been proposed in the context
of a converging sub--Keplerian flow within 50~\Rg~of the BH (e.g.,
Titarchuk \& Shrader 2002).  Turolla et al. (2002) have suggested
pair production as a means of extending the photon spectrum beyond the
$\sim$350 keV limit initially calculated for this model.

As noted above, Comptonization models are hard--pressed to explain the
origin of the energetic electrons.  As a further difficulty, a viable
model must also account for the QPOs in the SPL state.  This is
important since strong QPOs (see Fig.~\ref{spec-spl1} and
Fig.~\ref{spec-spl2}) are common in this state.  SPL models and X--ray
QPOs are discussed further in \S4.4 below.

\subsection{Intermediate states}

The four states described above capture the behavior observed for most
BHBs on many occasions.  However, other forms of complex behavior are
often seen, and these dispel the notion that every BHB observation can
be classified via these X--ray states.  In the following three
subsections, we consider observations that challenge or combine
elements of the four--state framework described above.

The {\it hard}--state energy spectra in Figures~\ref{spec-lh1} \&
\ref{spec-lh2} show little or no contribution from the accretion disk,
while the PDS exhibit a ``band--limited'' power continuum (i.e., a
flat power continuum at low frequencies that breaks to a steeper slope
between 0.1 and 10 Hz).  However, a band--limited power spectrum is
sometimes seen in combination with a stronger contribution from the
accretion disk.  This condition of a BHB has been interpreted as an
{\it intermediate} state that lies between the {\it hard} and TD
states (e.g., Mendez \& van der Klis 1997). We agree that the spectra
of GRO~J1655--40 shown in Figure~\ref{spec-lh1} do appear to represent
a legitimate example of the {\it intermediate} state in the sense of a
transition between the {\it hard} and TD states.

\begin{figure}
\epsfxsize28.5pc 
\epsfysize28.5pc 
\hspace*{0.08in}
\hspace*{-0.24in}
\epsfbox{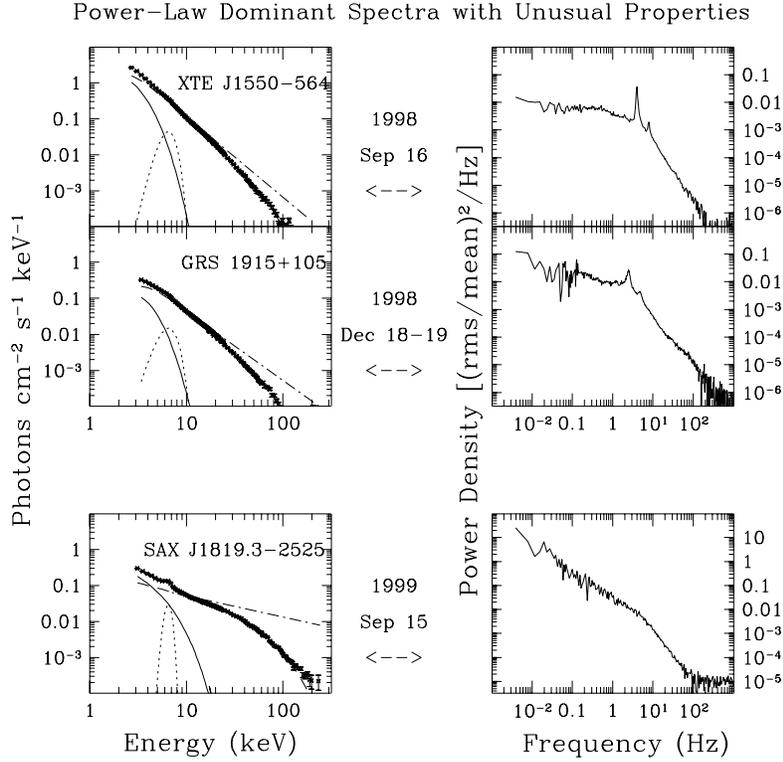}
\vspace*{-0.59in}
\caption{Unusual spectra of three BHBs. The observations of
XTE~J1550--564 and GRS~1915+105 show spectral properties of the SPL
state, but the PDS show a band--limited power continuum that is
customarily seen as a characteristic of the {\it hard} state, rather than the
SPL state.  In the bottom panel, the flares and rapid fluctuations
seen in SAX~J1819.3--2525 (V4641 Sgr) do not coincide with any of the
typical X--ray states of BHB systems.}
\label{spec-str}
\vspace*{-0.07in}
\end{figure}

Other cases seem to display a different type of intermediate or hybrid
emission state.  For example, dozens of observations of XTE~J1550--564
yielded energy spectra and QPOs that resemble the SPL state (Sobczak
et al. 2000b), but the PDS showed band--limited continuum power that
is reminiscent of the {\it intermediate} state described above (Homan
et al. 2001).  One example is shown in Figure~\ref{spec-str}, along
with a similar observation of GRS~1915+105.  These observations can be
described as SPL states with band--limited power continua. The
significance of this PDS shape is yet to be fully understood. It could
suggest an {\it intermediate} state linking the {\it hard} and SPL
states, a detail further supported by the fact that the disks in the
two systems appear cooler and larger than they do in the TD state.
However, such speculations may be ill--advised without considerations
of sensitive radio measurements.  We further note that the presence or
absence of band--limited power observed in individual SPL--state
observations of XTE~J1550--564 is closely coupled to the amplitudes
and phase lags of the associated QPOs (of types A, B, and C; Remillard
et al. 2002a). Finally, while the energy spectra of XTE~J1550--564 and
GRS~1915+105 in Figure~\ref{spec-str} are distinctly steep, there is
some ambiguity as to whether they are best modeled as a steep
power--law or as a flatter power--law with an unusually low cutoff
energy ($\sim 50$ keV). In Table~\ref{tab:spfit}, we show the spectral
parameters for the latter model, which is statistically preferred in
the case of GRS~1915+105.

We conclude that it is inappropriate to refer to both the observations
of XTE~J1550--564 (Fig.~\ref{spec-str}) and the very different
observations of GRO~J1655--40 (Fig.~\ref{spec-lh1}) as representing a
single BHB state, namely, the {\it intermediate} state.  On the other
hand, state transitions and hybrid emission properties are to be
expected and {\it X--ray spectra and PDS should be interpreted as {\it
intermediate} states when necessary, while specifying which states can
be combined to yield the the observed X--ray properties.}  In summary,
we describe the spectra of GRO~J1655--40 (Fig.~\ref{spec-lh1}) as
representing primarily a {\it hard} state or perhaps an {\it
intermediate} state between {\it hard} and TD.  On the other hand, the
spectra and PDS of XTE~J1550--564 and GRS~1915+105 considered here
appear to show an {\it intermediate} state related to the SPL and {\it
hard} states.  Since transitions between these latter two states are
not generally seen, this hybrid combination merits further scrutiny.

\subsection{X--ray states of Cygnus X--1 and GRS~1915+105}

In this section we briefly summarize the efforts to integrate the
behavior of two uncommon BHBs within the framework of the TD, {\it
hard}/radio jet, and SPL states.  We first return to the issue of Cyg
X--1 and the nature of its transitions to a soft state of high
intensity, which is unlike the canonical TD (HS) state.  As noted
earlier (\S4.2.1.1; \S4.3.1), contrary to expectations the 1996
soft--state spectrum of Cyg X--1 revealed a power--law spectrum,
rather than a TD spectrum (Zhang et al. 1997a; Frontera et al. 2001a;
Fig.~\ref{spec-spl2}).  Cyg X--1 has never been seen in the TD state,
and the transition from a hard X--ray spectrum to a soft one must be
seen as a transition from the {\it hard} state to the SPL state
(Gierlinski et al. 1999; Zdziarski et al. 2001).  However, this
SPL--like state is unusual in two respects: the absence of QPOs and
the relatively low temperature of the accretion disk
(Table~\ref{tab:spfit}).  Less surprising is the low luminosity of the
SPL state (Zhang et al. 1997b), since this is also seen in other
sources (e.g., XTE~J1550--564; Remillard et al. 2002a).  Whether the
SPL state in Cyg~X--1 requires a higher mass accretion rate than the
{\it hard} state is a matter of controversy (Zhang et al. 1997b;
Frontera et al. 2001a).

In the unique case of GRS~1915+105, the wildly varying X--ray light
curves indicate an imposing number of instability modes (Belloni et
al. 2000).  Nevertheless, within this complexity it is often possible
to identify the canonical states of a BHB (Muno et al. 1999; Belloni
et al. 2000). About half of the observations of GRS~1915+105 show
fairly steady X--ray flux (rms $<$ 15\% in 1 s bins at 2--30 keV), and
most of these intervals yield spectra and PDS that resemble either the
TD or {\it hard} states (Muno et al. 1999; Fender 2001; Klein--Wolt et
al. 2002). There are, however, some noteworthy anomalies encountered
while interpreting the behavior of GRS~1915+105 in terms of the
canonical X--ray states. First, the condition of steady radio and
X--ray emission extends to $L_{\rm x}~>~10^{38}$~erg~s$^{-1}$, which
is a factor $\sim 100$ higher than is observed for other BHBs in the
{\it hard} state. Secondly, the X--ray photon index ($\Gamma \sim
2.2$) is steeper than usual, although it remains flatter than the
index seen in in the SPL state ($\Gamma \ge 2.4$) or the hard tail of
the TD state ($\Gamma \approx 3$; see Table~\ref{tab:spfit}).  In short,
the spectral index for GRS~1915+105 in the {\it hard} state appears
shifted to somewhat high value compared to other BHBs.

Overall, the spectral and temporal properties of Cyg X--1 and
GRS~1915+105 are best integrated into the standard description of BH
X--ray states if we relax the assumptions regarding the relative or
absolute luminosity ranges that are appropriate for the various
states. As we will see in \S4.5.1, the spectral evolution of
XTE~J1550--564 motivates a similar conclusion, since the luminosity in
the SPL state can lie well below that of the TD state.  Undoubtedly,
there is an overall correlation between spectral states and luminosity
intervals in accreting BHB systems.  However, {\it the canonical X--ray
states are most usefully defined in terms of the properties of the
energy spectrum and PDS, rather than in terms of luminosity}.

\begin{rotation}
{\fontsize{7}{9}\selectfont
\caption{Spectral Fit Parameters}
\label{tab:spfit}
\begin{tabular*}{38pc}{@{}l\x c\x c\x c\x c\x c\x c\x c\x c\x c\x c\x c\x
c\x c\x l@{}}
\hline\hline
{X--ray} & {$N_H$} & {$T_{DBB}$} &  {$\pm$} &  {$N_{DBB}$~~} & {$\pm$~~} &
{$\Gamma_{PL}$~~} &  {$\pm$~~~} & {$N_{PL}$~~~} & {$\pm$} & {Fe} &  {$N_{Fe}$~~}
&   {$\pm$~~~} & {$\chi_{\nu}^2$} & {~~additional details~~~~~~~~~~~~}\\
name      & ($10^{22}$) & (keV) & (keV) &       &      &    &    &     &
       & FWHM &       &       &           &  \\ 
\hline\hline
\hspace*{0.10in}TD State: Figs.~\ref{spec-td1} \& \ref{spec-td2} \\
4U 1543-475      &  0.3 &  1.01 & 0.02 & 7419. & 165.~~& 2.57~~& 0.02~~~& 5.42~~~& 0.21  & 0.61 & .0479~~& .0031~~~&~3.62~&~~feature at 4.4 keV \\
XTE J1550-564    &  2.0 &  1.12 & 0.03 & 3289. &  74.~~& 4.76~~& 0.04~~~& 152.~~~&  17.  & ...  &  ...  & ...   &~0.98~&~~smedge at 9.2 keV \\
GRO J1655-40     &  0.9 &  1.16 & 0.03 & 1559. &  21.~~& 2.85~~& 0.23~~~& 1.01~~~& 0.65  & ...  &  ...  & ...   &~2.00~&~~smedge at 8.0 keV \\
GX 339-4         &  0.2 &  0.71 & 0.03 & 2520. &  62.~~& 2.02~~& 0.04~~~& 0.08~~~& 0.01  & 1.05 & .0032~~& .0003~~~&~1.45~& \\
GRS1915+105      &  6.0 &  2.19 & 0.04 &   62. &   5. & 3.46~~& 0.02~~~& 33.4~~~& 1.61  & ...  &  ...  & ...   &~3.13~&~~smedge at 6.7 keV \\
4U 1630-47       & 11.0 &  1.33 & 0.03 &  315. &   7. & 3.75~~& 0.03~~~& 17.4~~~& 1.40  &  ... &   ... &  ...  &~1.06~&~~break at 20.8 keV to $\Gamma=1.9$ \\
GRS 1739-278     &  3.0 &  0.95 & 0.04 &  972. &  23.~~& 2.65~~& 0.15~~~& .210~~~& 0.008 & 1.11 & .0068~~& .0008~~~& 1.42 & \\
XTE J1748-288    & 10.4 & 1.79 & 0.02 &  42.4  &  2.1~~& 2.60~~& 0.02~~~& 14.6~~~& 0.4  &  ... &  ...  &  ...  & 1.18 & \\
XTE J1755-324    &  0.2 &  0.75 & 0.08 & 1486. & 133.~~& 2.40~~& 0.15~~~& 0.11~~~& 0.04  & ...  &  ...  & ...   & 1.78 & \\
XTE J2012+381    &  0.8 &  0.85 & 0.05 & 1176. &  56.~~& 2.06~~& 0.04~~~& 0.16~~~& 0.015 & ...  &  ...  & ...   & 1.32 & \\
\hspace*{0.10in}{\it Hard} State: Figs.~\ref{spec-lh1} \& \ref{spec-lh2} \\
4U 1543-475      &  0.3 & 0.38 & 0.07  & 645. & 1338.~~& 1.67~~& 0.02~~~& 0.041~~~& 0.001 & ... & ... & & 1.57 & \\
XTE J1550-564    &  2.0 &  ...  & ...  &  ....  & ... & 1.70~~& 0.10~~~& 0.108~~~& 0.021 & ...  &  ...  & ...   & 1.13 & \\
GRO J1655-40     &  0.9 &  0.77 & 0.02 &  228.  & 37.~~& 1.93~~& 0.02~~~& 0.571~~~& 0.021 & 1.00 & .0065~~& .0006~~~& 1.83 & \\
GX 339-4         &  0.2 &  ...  & ...  &  ....  & ... & 1.75~~& 0.02~~~& 0.168~~~& 0.028 & 0.90 & .0013~~& .0003~~~& 0.98 &~~plus reflection \\
GRS1915+105      &  6.0 &  ...  & ...  &  ....  & ... & 2.11~~& 0.02~~~& 0.231~~~& 0.043 & 0.91 & .0458~~& .0003~~~& 2.39 &~~plus reflection \\
XTE J1118+480    &  0.01 &  ...  & ...  &  ....  & ... & 1.72~~& 0.04~~~& 0.267~~~& 0.024 & ...  & ... & ...   & 1.23 & \\
GS 1354-644      &  0.7  &  ...  & ...  &  ....  & ... & 1.48~~& 0.09~~~& 0.470~~~& 0.032 & 0.1  & .0008~~& .0002~~~& 1.15 &~~plus reflection \\
XTE J1748-288    & 10.4  &  0.48 & 0.05 &  5302. & 479.~~& 1.88~~& 0.09~~~& 0.293~~~& 0.065 & 0.66 & .0045~~& .0003~~~& 1.65 & \\
GRS 1758-258     &  1.0  &  ...  & ...  &  ....  & ...  & 1.67~~& 0.07~~~& 0.053~~~& 0.010 & 0.36 & .0004~~& .0001~~~& 1.84 & \\
Cyg X-1          &   0.5 &  ...  & ...  &  ....  & ...  & 1.68~~& 0.07~~~& 0.446~~~& 0.025 & 1.44 & .0206~~& .0018~~~& 3.40 &~~plus reflection \\
\hspace*{0.10in}SPL State: Figs.~\ref{spec-spl1} \& \ref{spec-spl2} \\
4U 1543-47       & 0.3  & 0.93 & 0.07 &  3137. &  138.~~& 2.47~~& 0.02~~~& 6.85~~~& 0.21 & 0.82 & .0347~~& .0034~~~& 1.90 & \\
XTE J1550-564    &  2.0 & 3.31 & 0.20 &   7.76 &  0.70~~& 2.82~~& 0.05~~~& 200.~~~& 1.5  & 1.30 & .2136~~& .0314~~~& 2.16 &~~smedge at 8.5 keV \\
GRO J1655-40     &  0.9 & 2.22 & 0.20 &   9.89 &  1.6~~&   2.65~~& 0.05~~~& 75.3~~~& 1.1  & 1.32 & .2321~~& .0157~~~& 4.45  & \\
GX 339-4         &  0.2 & 0.89 & 0.08 &  1917. & 109.~~&  2.42~~& 0.02~~~& 2.34~~~& 0.08 & 0.97 & .0178~~& .0017~~~& 1.39 & \\
GRS 1915+105     &  6.0 & 1.19 & 0.07 &  115.  &  31.~~&  2.62~~& 0.08~~~& 28.5~~~& 0.6  & 0.90 & .0396~~& .0065~~~& 4.13 & \\
4U 1630-47       & 11.0 & 1.73 & 0.02 &  46.0 &   2.4~~&  2.65~~& 0.02~~~& 17.0~~~& 0.4  & ...  &  ...  &  ...  &  1.10 & \\
GRS 1739-278     &  3.0 & 1.01 & 0.06 &  1116. &  38.~~&  2.61~~& 0.03~~~& 2.95~~~& 0.19 & 1.53 & .0341~~& .0021~~~& 0.94 & \\
XTE J1748-288    & 10.4 & 1.36 & 0.02 &  210.  &  11.~~&  2.92~~& 0.02~~~& 26.2~~~& 1.2  &  ... &  ...  &  ...  & 0.96 & \\
XTE J1859+226    &  0.5 & 1.03 & 0.02 &  1164. &  91.~~&  2.55~~& 0.08~~~& 14.5~~~& 0.31 & 1.33 & .0426~~& .0060~~~& 1.36 & \\
Cyg X-1          &  0.5 & 0.49 & 0.03 &  55708.~~& 2962.~~& 2.68~~&
0.03~~~& 7.65~~~& 0.37 & 0.73 & .0270 & .0016 & 1.78 &~~breaks to $\Gamma=1.83$ above 12 keV \\
\hspace*{0.10in}Unusual Spectra: Fig.~\ref{spec-str} \\
XTE J1550-564    &  2.0 & 0.74 & 0.02 & 6932. & 562.~~&  2.24~~& 0.02~~~& 23.0~~~& 0.87 & 1.03 & .121~~& .008~~~& 1.34 &~~p.l. cutoff energy 51.7 keV \\
GRS 1915+105     &  6.0 & 0.88 & 0.03 &  775. & 156.~~&  1.91~~& 0.02~~~& 4.51~~~& 0.18 & 1.28 & .053~~& .004~~~& 1.45 &~~p.l. cutoff energy 50.0 keV \\
SAX J1819.3-2525 &  0.3 & 1.63 & 0.06 &  38.  &   6.~~&  0.59~~& 0.03~~~& 0.25~~~& 0.02 & 0.47 & .038~~& .003~~~& 1.46 &~~p.l. cutoff energy 39.3 keV \\
\hline\hline
\end{tabular*}}
\end{rotation}

\subsection{Anomalous behavior of SAX~J1819.3--2525}

Finally, we consider whether BHBs exhibit characteristics that fall
entirely outside the X--ray states considered thus far.  The most
challenging case may be the BHB SAX~J1819.3--2525 (V4641 Sgr;
Fig.~\ref{asm_bh}). The PCA observations of 1999 September 15.9 show
the source in a unique flaring state (Wijnands \& van der Klis
2000). The spectral hardness ratio remains remarkably constant
throughout the brightest $\sim$500~s of this observation, despite the
dramatic intensity fluctuations.  The spectrum and PDS for this
central time interval are shown in Figure~\ref{spec-str}. The spectral
analysis reveals a hot accretion disk and a broad Fe line, while the
photon index is extraordinarily hard: $\Gamma = 0.60 \pm 0.03$ with a
cutoff energy of 39~keV (Table~\ref{tab:spfit}).  This source is also
distinguished as a new prototype for a ``fast X--ray nova'' (Wijnands
\& van der Klis 2000) because it exhibited a 12~Crab flare that
appeared and decayed in less than 1 day (Fig.~\ref{asm_bh}).  The
multifrequency spectrum near X--ray maximum has been interpreted in
terms of super--Eddington accretion with the binary immersed in an
extended envelope (Revnivtsev et al. 2002).


\section{Fast temporal variations: QPOs and broad power peaks }

As shown in the preceding sections on X--ray states (\S4.3.6--8), QPOs
are prevalent in the SPL state, and they are sometimes seen in the
{\it hard} state when thermal emission from the disk contributes some
flux above 2~keV (e.g., GRO~J1655-40 in Fig. 4.11).  In this section
we briefly consider the QPOs of BHBs and BHCs in greater detail.  For
references on PDS computation, defining QPOs, and QPO characteristics
of NS systems, see Chapter~2.  We supplement this work by discussing
X--ray timing results for BHBs.  Following van der Klis, we define
QPOs as features (usually modeled as a Lorentzian function) in the PDS
that have coherence parameter $Q = \nu / \Delta \nu > 2$
(FWHM). Features with significantly lower $Q$ values are regarded as
``broad power peaks'' and are discussed separately.

\subsection{Low frequency QPOs and radiation mechanisms}

The X--ray PDS of many BH transients display low frequency QPOs
(LFQPOs) roughly in the range 0.1 to 30 Hz. The significance of these
oscillations can be summarized as follows.

(1) LFQPOs are almost always seen during the SPL state. They can be
exceedingly strong (see Figs.~\ref{spec-spl1}--\ref{spec-str}) with
rms amplitudes (expressed as a fraction of the mean count rate) as
high as $r > 0.15$ for sources such as GRS~1915+105 (Morgan et
al. 1997) and XTE~J1550--564 (Sobczak et al. 2000a). More generally,
they are seen with $0.03 < r < 0.15$ whenever the steep power law
contributes more than 20\% of the flux at 2--20 keV (Sobczak et
al. 2000a).  LFQPOs have been observed at energies above 60~keV
(Tomsick \& Kaaret 2001).

(2) In several sources, the LFQPO frequency is correlated with the
total disk flux (but not with temperature or inner disk radius;
Sobczak et al. 2000a; Muno et al. 1999; Trudolyubov et al. 1999).
This behavior, in combination with the role of the steep power law
mentioned directly above, suggests that LFQPOs may provide a vital
clue to the mechanism that couples the thermal and SPL components.

(3) LFQPOs can be quasi--stable features that persist for days or
weeks.  For example, in GRS~1915+105 
QPOs at 2.0--4.5 Hz persisted for 6 months
during late 1996 and early 1997 (Muno et al. 2001).

(4) In a general sense, it can be argued that oscillations as distinct
and strong as these QPOs (often with $Q > 10$), represent global
requirements for an organized emitting region. For example, in the
context of models in which thermal radiation originates from MHD
instabilities, one cannot accept the common picture of numerous and
independent magnetic cells distributed throughout the inner disk.

The effort to tie LFQPOs to the geometry and flow of accreting gas is
complicated by the fact that LFQPO frequencies are much lower than the
Keplerian frequencies for orbits in the inner accretion disk. For
example, for a BH mass of 10~\msun, an orbital frequency near 3 Hz
coincides with a disk radius near 100 $R_g$, while the expected radius
for maximum X--ray emission lies in the range 1--10 $R_g$ (depending
on the value of the BH spin parameter). For the strongest QPOs in
GRS~1915+105, the individual oscillations were tracked to determine
the origin of frequency drifts and to measure the average
``QPO--folded'' oscillation profile (Morgan et al. 1997). The results
show a random walk in QPO phase and a nearly sinusoidal waveform. The
ramifications of these results for QPO models remain uncertain.

There are now a large number of proposed LFQPO mechanisms in the
literature, and we mention only a few examples here. The models are
driven by the need to account for both the QPO frequency and the
fact that
the oscillations are strongest at photon energies above 6 keV,
i.e., where only the power--law component contributes substantially 
to the X--ray spectrum.
The models include global disk
oscillations (Titarchuk \& Osherovich 2000), radial oscillations of
accretion structures such as shock fronts (Chakrabarti \& Manickam
2000), and oscillations in a transition layer between the disk and a
hotter Comptonizing region (Nobili et al. 2000). Another alternative,
known as the ``accretion--ejection instability model,'' invokes spiral
waves in a magnetized disk (Tagger \& Pellat 1999) with a transfer of
energy out to the radius where material corotates with the spiral
wave. This model thereby combines magnetic instabilities with
Keplerian motion to explain the observed QPO amplitudes and stability.

Further analyses have revealed phase lags associated with LFQPOs and
their harmonics. The analysis technique uses Fourier cross spectra to
measure both the phase lags and the coherence parameter ({\it versus}
frequency) between different X--ray energy bands, e.g., 2--6 {\it vs.}
13--30 keV.  Unexpectedly, both positive and negative phase lags 
have been found (Wijnands et al. 1999; 
Cui et al. 2000b; Reig et al. 2000; Muno
et al. 2001), and efforts have been made to classify LFQPOs by phase
lag properties.  The expansion of LFQPO subtypes may not be widely
viewed as a welcome development.  Nevertheless, it has been shown in
the case of XTE~J1550--564 that the properties of the phase lags
clarify how LFQPO parameters correlate with both the accretion--disk
and high--frequency QPO parameters (Remillard et al. 2002a).

\subsection{Broad power peaks and comparisons of BH and NS 
systems}

The study of broad features in the PDS has led recently to important
developments.  In many NS systems and in the {\it hard} state of BHBs,
the PDS can be decomposed into a set of four or five broad power peaks
(Nowak 2000; Belloni et al. 2002), generally with $0.5 < Q < 1.0$.
The evolution of these features has been linked to major behavioral
changes in Cyg X--1.  For example, the disappearance of the third
broad power peak occurred just at the time Cyg X--1 left the {\it
hard} state and its steady radio jet was quenched (Pottschmidt et
al. 2003).

Broad PDS features are also involved in renewed efforts to contrast
accreting BHB and NS systems via their variability characteristics.
It has been proposed that the observed high--frequency limit of the
power continuum provides a means to distinguish accreting BH and NS
systems (Sunyaev \& Revnivtsev 2000), since only the latter exhibit
intensity variations above 500 Hz.

Finally, some studies have used both QPOs and broad power peaks to
examine the relationship between low-- and high--frequency features
and to make comparisons between different NS subclasses and
BHBs. There have been claims of a unified variability scheme that
encompasses all X--ray binary types (Psaltis et al. 1999; Belloni et
al. 2002).  The bottom line of this scheme is that all of the
oscillations must originate in the accretion disk. However, important
aspects of this work remain controversial, particularly the handling
of BH HFQPOs and their association with the lower kHz QPO observed for
NSs (Remillard et al. 2002a).

\subsection{High--frequency QPOs and general relativity}

\begin{figure}
\epsfxsize34pc 
\epsfysize34pc 
\hspace*{-0.42in}
\hspace*{-0.30in}
\epsfbox{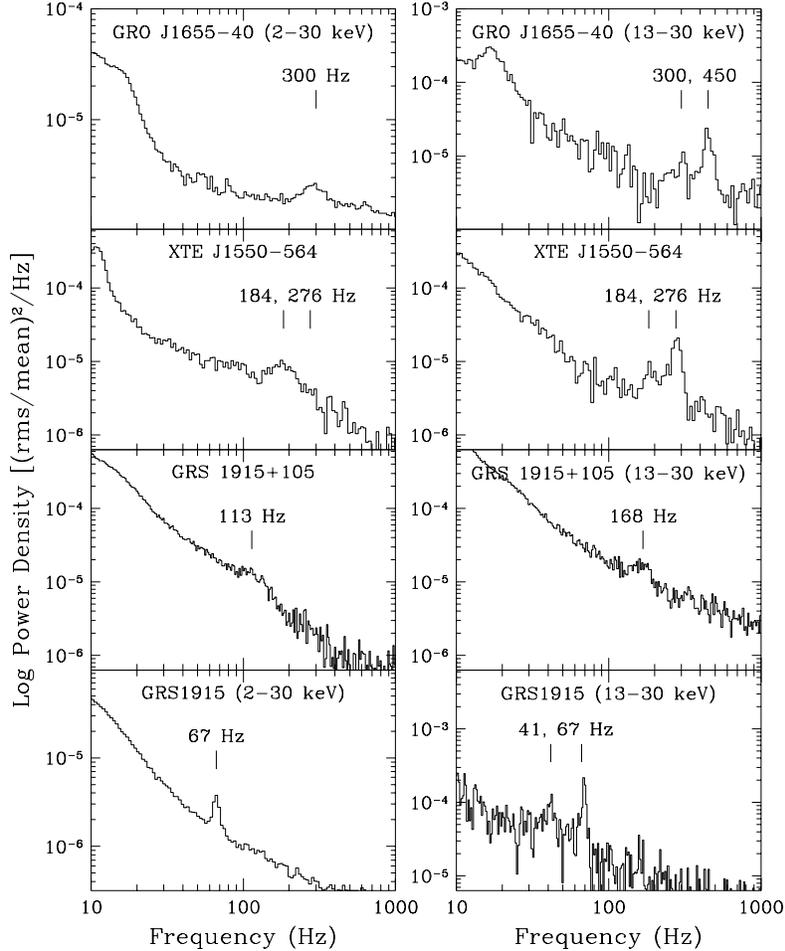}
\vspace*{-0.08in}
\caption{Four pairs of HFQPOs observed in three black--hole binary
systems.  The energy band is 6--30 keV unless otherwise
indicated. These usually subtle oscillations are only visible during a
fraction of the observations for each source.}
\label{hfqpo_pairs}
\vspace*{-0.07in}
\end{figure}

The topic of high--frequency QPOs (HFQPOs) in BHBs (40--450 Hz)
continues to evolve in the {\it RXTE} era. These transient QPOs have
been detected in seven sources (4 BHBs and 3 BHCs). HFQPOs have rms
amplitudes that are generally $\sim$1--3\% of the mean count rate in a
given energy band. Remarkably, three sources exhibit pairs of QPOs
that have commensurate frequencies in a 3:2 ratio (Remillard et
al. 2002b; Remillard et al. 2003b; Table~\ref{tab:bhb2}). As shown in
Figure~\ref{hfqpo_pairs}, GRO~J1655-40 and XTE~J1550-564 each exhibit
a single such pair of frequencies.  GRS~1915+105, on the other hand,
shows two pairs of HFQPOs; the complete set of four QPO pairs is shown
in Figure~\ref{hfqpo_pairs}.  In addition (see references in
Tables~\ref{tab:bhb2} \& \ref{tab:bhc}), single--component HFQPOs have
been observed in 4U1630--47 (184~Hz), XTE~J1859+226 (190~Hz),
XTE~J1650--500 (250~Hz), and H~1743--322 (240~Hz).  Their profiles are
similar to the 300 Hz QPO for GRO~J1655--40 shown in the top left
panel of Figure~\ref{hfqpo_pairs}.  HFQPOs occur in the SPL state,
except for the 67 Hz QPO in GRS~1915+105 (see Fig.~\ref{hfqpo_pairs})
which appears in the TD state, especially when $L_{x} >~10^{38}$
erg~s$^{-1}$.

\begin{figure}
\epsfxsize19.5pc 
\epsfysize19.5pc 
\hspace*{-0.50in}
\hspace*{0.05in}
\epsfbox{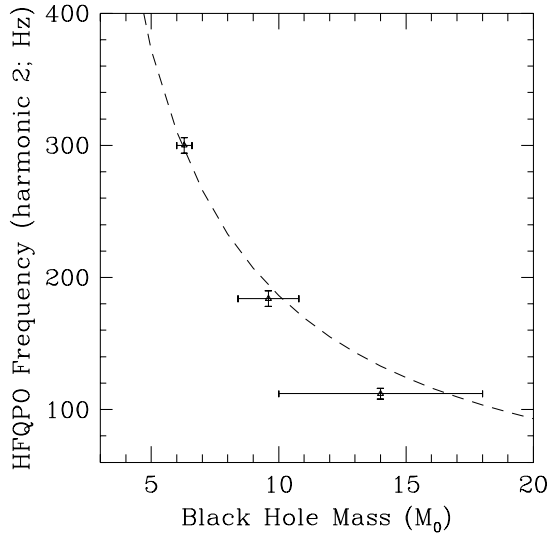}
\vspace*{-0.43in}
\caption{Relationship between HFQPO frequency and BH mass for
XTE~J1550--564, GRO~J1655--40, and GRS~1915+105.  These three systems
display a pair of HFQPOs with a 3:2 frequency ratio.  The frequencies
are plotted for the stronger QPO that represents $2 \times \nu_0$. The
fundamental is generally not seen in the power spectra.  The dashed
line shows a relation, $\nu_0$ (Hz) = 931 ($M$/\msun)$^{-1}$, 
that fits these data.}
\label{hfqpo_m}
\end{figure}

The preponderance of evidence indicates that HFQPOs do not shift
freely in frequency in response to luminosity changes, as do the kHz
QPOs in NS systems (see Ch. 2).  Instead, they appear to exhibit an
``X--ray voiceprint''.  That is, the QPOs occur in harmonics of an
unseen fundamental frequency which has a unique value for each BH.
For the three cases that show 3:2 frequency pairs, the relationship
between the HFQPO frequencies {\it vs.} BH mass scales as
$M_{1}^{-1}$. This relationship is shown in Figure~\ref{hfqpo_m}, where
we have plotted the frequency of the stronger feature (i.e. $2~\times
\nu_0$), since the fundamental is generally not seen. These results
offer strong encouragement for seeking interpretations of BH HFQPOs
via GR theory, since each type of GR disk oscillation under strong
gravity varies as $M_{1}^{-1}$, assuming the sampled BHs have similar
values of the dimensionless spin parameter ($a_*$).

These commensurate frequencies can be seen as strong support for the
idea that HFQPOs may represent some type of resonance phenomenon
involving oscillations describable by GR, as originally proposed by
Abramowicz \& Kluzniak (2001).  Resonances in some form may be
applicable to both BH and NS systems (Abramowicz et al. 2003).  We
note that the 3:2 harmonic pattern cannot be attributed to a distorted
sine wave with harmonic content because the individual detections (in
a given energy band, on a given day) generally appear as a single peak
in the PDS, and the presence of a pair of commensurate frequencies is
recognized only when the ensemble of results is examined.

Coordinate frequencies and their differences (i.e., beat frequencies)
in GR were proposed earlier to explain some of the X-ray QPOs from
both NSs and BHs (Stella et al. 1999); however, this work did not
treat the commensurate frequencies of interest here, which were 
discovered subsequently in three BHB systems.  In the
resonance hypothesis (Abramowicz \& Kluzniak 2001), these harmonic
frequencies are discussed in terms of accretion blobs following
perturbed orbits in the inner accretion disk.  Unlike Newtonian
gravity, GR predicts independent oscillation frequencies for each
spatial coordinate for orbits around a rotating compact object, as
seen from the rest frame of a distant observer.  Over the range of 
radii in the accretion disk where X--rays are expected to originate
in the standard disk model, pairs of GR coordinate frequencies have 
varying, non--integral ratios.  Therefore, the
discovery of HFQPOs with commensurate frequencies can be seen to
suggest enhanced emissivity at a particular radius where a pair of
coordinate frequencies are in some type of resonance.  Unlike the
azimuthal and polar coordinate frequencies, the radial coordinate
frequency reaches a maximum value and then falls to zero as the radius
decreases toward $R_{\rm ISCO}$ (Kato 2001; Merloni et al. 2001). This
ensures the possibility of commensurate coordinate frequencies
somewhere in the inner disk. For example, there is a wide range in the
dimensionless spin parameter, $a_*$, where one can find a particular
radius that corresponds to a 2:1, 3:1, or 3:2 ratio in the orbital and
radial coordinate frequencies.  A resonance between the polar and
radial coordinate frequencies is also possible.  In the resonance
scenario, linear perturbations may grow at these radii, ultimately
producing X--ray oscillations that represent some combination of the
individual resonance frequencies, their sum, or their
difference. However, there remain serious uncertainties as to whether
such structures could overcome the severe damping forces and emit
X--rays with sufficient amplitude and coherence to produce the
observed HFQPOs (Markovic \& Lamb 1998).

Models for ``diskoseismic'' oscillations adopt a more global view of
the inner disk as a GR resonance cavity (Kato \& Fukue 1980; Wagoner
1999). This paradigm has certain attractions for explaining HFQPOs,
but integral harmonics are not predicted for the three types of
diskoseismic modes derived for adiabatic perturbations in a thin
accretion disk.  Clearly, there is a need to investigate further the
possibility of resonances within the paradigm of diskoseismology.
Other models must be considered as well, e.g. the p-mode oscillations 
of an accretion torus surrounding a black hole (Rezzolla et al. 2003).

It has been argued by Strohmayer (2001a) that HFQPO frequencies are
sufficiently high that they require substantial BH spin.  For example,
in the case of GRO~1655--40 the 450~Hz frequency exceeds the maximum
orbital frequency ($\nu_{\phi}$) at the ISCO around a Schwarzschild BH
(i.e., $a_* = 0$; \S4.1.5) of mass $M_{1} = 6.3\pm0.5$~\msun~(Greene
et al.  2001).  {\it If} the maximum Keplerian frequency is the
highest frequency at which a QPO can be seen, then the results for
GRO~J1655--40 require a Kerr BH with prograde spin, for example, $a_*
> 0.15$. However, the conclusion that spin is required may not be
valid if the QPO represents the sum of two beating frequencies.  On
the other hand, even higher values of the spin parameter may be
required if the QPO represents either resonant coordinate frequencies
($a_* > 0.3$; Remillard et al. 2002b) or diskoseismic oscillations
($a_* > 0.9$; Wagoner et al. 2001).

Accurate and sensitive measurements of X--ray HFQPO frequencies may
lead to a determination of the GR mechanism that is responsible for
these oscillations, and ultimately to secure measurements of the spin
parameter $a_*$ for a number of BHs.  These spin measurements would be
very valuable in assessing the role of BH rotation in the production
of jets (Blandford \& Znajek 1977).  This is especially true since the
three systems with paired frequencies have a history of relativistic
mass ejections during some (but not all) of their outbursts.

\section{Energetics and key variables determining BHB radiation}

\subsection{Division of spectral energy: disk and power--law components}

\begin{figure}
\epsfxsize25pc 
\epsfysize25pc 
\hspace*{-0.28in}
\hspace*{0.12in}
\epsfbox{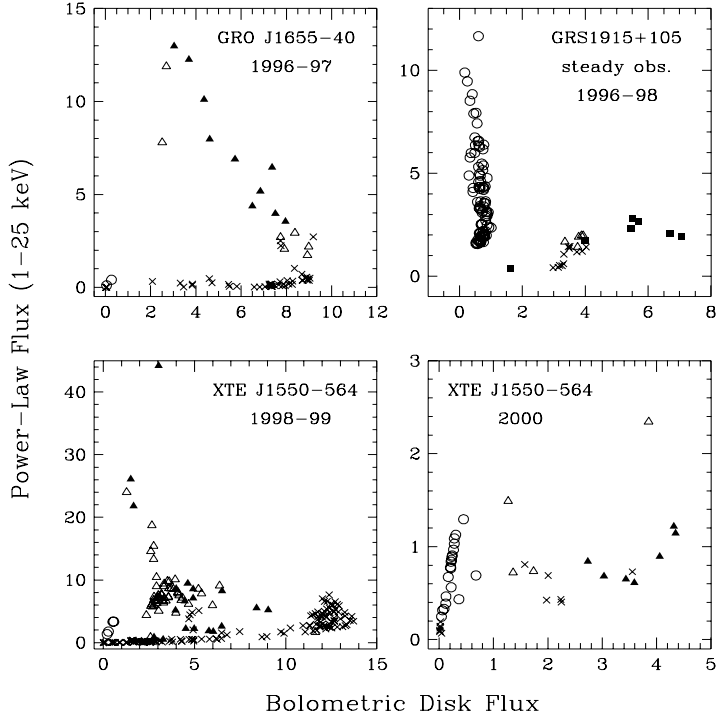}
\vspace*{-0.16in}
\caption{Radiation energy division between the accretion disk and the
power--law component.  The symbol types denote the X--ray state as TD
(x), {\it hard} ($\circ$), or SPL($\Delta$).  Furthermore, in the SPL
state, an open triangle is used when there is only an LFQPO, while a
filled triangle denoted the additional presence of an HFQPO.  Finally,
the TD--like observations that show 67 Hz QPOs in GRS~1915+105 are
shown with a filled square.  All fluxes are in units of
$10^{-8}$~\ergcm and corrected for absorption.}
\label{fig:flux3}
\vspace*{-0.07in}
\end{figure}

A revealing view of the behavior of a BHB over the course of its
entire outburst cycle can be obtained by plotting the flux in the
power--law component vs. the flux from the accretion disk.  Four such
plots are shown in Figure~\ref{fig:flux3}, where the various plotting
symbols identify the emission states described in \S4.3.  The figure
shows the flux diagrams for 3 BHBs, while considering two of the
outbursts of XTE J1550--564 (Muno et al. 1999; Remillard et al. 2001;
Remillard et al. 2002b).  It has been shown that the emerging patterns
in these flux diagrams are largely unaffected by the choice of
integration limits, e.g., whether the energy measurements are computed
in terms of bolometric flux or the integrated flux within the 2--25
keV PCA bandpass (Remillard et al. 2002a).

The TD points (``x'' symbol; Fig.~\ref{fig:flux3}) for GRO~J1655--40
and XTE~J1550--564 (1998--99) appear well organized; they can be
described as horizontal tracks in which accretion energy is freely
converted to thermal radiation from the accretion disk.  These tracks
correspond to the standard accretion disk model (Shakura \& Sunyaev
1973), and they may also convey moderate Comptonization effects
expected from MHD turbulence (e.g., Hawley \& Krolik 2001) as these
tracks curve upward at high luminosity.  The TD points at the highest
flux levels correspond to Eddington luminosities of 0.2 for
GRO~J1655--40, 0.6 for XTE~J1550--564 (1998--99), and 0.4 for
GRS~1915+105, using the values for mass and distance given in
Tables~\ref{tab:bhb1} \& \ref{tab:bhb2}.

Observations in the {\it hard} state (i.e., a dominant power--law
component with $1.4 < \Gamma < 2.2$) are plotted as circles in
Figure~\ref{fig:flux3}.  These points form vertical tracks with only
minor flux contributions from the disk. The {\it hard}--state points
coincide with fairly steady radio emission for GRS1915+105 (Muno et
al. 2001) and XTE~J1550--564 in 2000 (Corbel et al. 2001), as expected
for this state.

Observations in the SPL state (defined by a power--law index $\Gamma >
2.4$ and by QPOs) are plotted as triangles in Figure~\ref{fig:flux3};
a solid symbol is used when there is an additional detection of an
HFQPO above 100 Hz. The wide diversity in the relative contributions
from the disk and power--law components within the SPL state are
especially apparent for XTE~J1550--564.  It is these results, combined
with the wide range in the luminosities of the TD and {\it hard}
tracks shown in Figure~\ref{fig:flux3}, that have caused us to abandon
the luminosity requirements in defining the X--ray states of BHBs
(\S4.3). Finally, detections of the 67 Hz QPO in GRS~1915+105 are
shown as filled squares. They appear to extend the TD branch out to
0.6 $L_{Edd}$.

\vspace*{-0.10in}

\subsection{Key variables determining BHB radiation}

Prior to the {\it RXTE} era, it was thought that the X--ray states
primarily represent a simple progression in luminosity, and that the
emission properties depend only on the accretion rate and the BH mass
(e.g., Tanaka \& Lewin 1995).  The behavioral complexity of sources
such as XTE~J1550--564 (Fig.~\ref{fig:flux3}) have challenged this
viewpoint (Homan et al 2001; Remillard et al. 2002b) because it now
appears that any of the three active states of accretion may occur at
a given luminosity.

Finally, what other parameters must be considered in the theory
of BH accretion?  It is clear that BH spin and the angle between the
spin and disk axes must also be considered in a complete
theory of BH accretion, but these would not help to explain the types
of rapid state transitions seen in many systems.  As noted in several
previous sections, MHD instabilities are widely invoked to explain
nonthermal radiation, and magnetic fields are additionally expected to
play a leading role in the formation and collimation of jets.  It then
seems relevant to question whether a parameter of magnetism, such as
the ratio of magnetic to gas pressure, should be considered a key
variable in accretion physics. The global magnetic field geometry may
also play an essential role in the formation of some X-ray
states.  The continued development of 3--D MHD simulations are
expected to be very fruitful in gaining a deeper understanding of the
radiation emitted by black hole binaries.

\vspace*{-0.10in}

\section{Concluding remarks}

As reviewed in \S4.4.3, high frequency QPOs at 40--450~Hz have been
observed with {\it RXTE} for four BHBs and three BHCs.  Furthermore,
three of these sources show harmonic (3:2) pairs of frequencies that
scale as $M_{BH}^{-1}$.  The models for these QPOs (e.g., orbital
resonance and diskoseismic oscillations) invoke strong--field GR
effects in the inner accretion disk, and they depend on both the mass
and spin of the BH.  On the other hand, studies of broadened Fe
K$\alpha$ emission lines (\S4.2.3), which can also reveal the
conditions in the very inner accretion disk, may prove as revealing as
timing studies.  The line photons that reach a distant observer are
gravitationally redshifted and Doppler and transverse--Doppler
shifted.  Encoded in the line profile are the mass and spin of the BH.
The most provocative result has been obtained for the BHC
XTE~J1650--500 (\S4.2.3), where the line profile suggests the presence
of an extreme Kerr BH.

The behavior of the massive compact objects reviewed herein supports
the view that they are bona fide BHs, which are described by GR and
were formed by the complete gravitational collapse of matter.
Our challenge is to prove that this conclusion is correct by making
clean quantitative measurements of relativistic effects in the strong
gravitational fields near these objects.  At present, no one can say
which future mission can best help us meet this challenge: LISA,
MAXIM, Constellation--X, or an X--ray Timing Observatory (XTO).
Perhaps all of the approaches will be required. In any case, an XTO
with effective area and telemetry ten times that of {\it RXTE} and
with improved energy resolution, would be a powerful probe of physics
near the event horizon.  Consider that {\it RXTE} with just 1.6 times
the effective area of {\it Ginga}, broke through to discover kHz QPOs
in NSs, a discovery that led quickly to hard constraints on dense
matter.

\vspace*{-0.10in}

\section*{Acknowledgments}

We thank Jon Miller, Mike Muno, David Smith, Jean Swank, John Tomsick,
and Rudy Wijnands for their help in assembling the catalogue of
candidate BHBs presented in Table~\ref{tab:bhc}, and Keith Arnaud,
George Field, Mike Garcia, Aya Kubota, Kazuo Makishima, and Mike
Revnivtsev for valuable comments. We are indebted to Andy Fabian, Jon
Miller, Ramesh Narayan, Jerry Orosz and Andrej Zdziarski for a careful
reading of portions of the manuscript and their valuable comments.  We
are especially grateful to Jeroen Homan and John Tomsick for their
detailed comments on a near--final draft of the manuscript.  We thank
Ann Esin for supplying Figure~\ref{adaf_esin}, Jon Miller for
supplying Figure~\ref{miller}, John Tomsick and Emrah Kalemci for the
use of unpublished data on 4U1543--47, John Huchra for advice on the
distance to the LMC, and Suresh Kumar for his assistance in
typesetting the manuscript.  This work was supported in part by NASA
under Grants NAG5--10813 and by the NASA contract to M.I.T. for
support of {\it RXTE}.

{}

\end{document}